\documentclass{optica-article}

\journal{opticajournal} 

\articletype{Research Article}

\usepackage{lineno}

\usepackage{xcolor}
\newcommand{\red}[1]{{\normalcolor#1}}   
\newcommand{\blue}[1]{{\normalcolor#1}}   

\usepackage{amsmath}

\begin{document}

\title{Generalized Quantum Geometric Tensor in a Non-Hermitian Exciton-Polariton System}

\author{Y.-M. Robin Hu$^*$, Elena A. Ostrovskaya$^\dagger$, and Eliezer Estrecho$^\ddagger$}

\address{\authormark{1}ARC Centre of Excellence in Future Low-Energy Electronics Technologies and Department of Quantum Science and Technology, Research School of Physics, The Australian National University, Canberra 2601, Australia}

\email{\authormark{*}yow-ming.hu@anu.edu.au, \authormark{$\dagger$}elena.ostrovskaya@anu.edu.au, \authormark{$\ddagger$}eliezer.estrecho@anu.edu.au} 


\begin{abstract*} 
In this work, we review different generalizations of the quantum geometric tensor (QGT) in two-band non-Hermitian systems and propose a protocol for measuring them in experiments. We present the generalized QGT components, i.e. the quantum metric and Berry curvature, for a non-Hermitian hybrid photonic (exciton-polariton) system and show that the generalized non-Hermitian QGT can be constructed from experimental observables. In particular, we extend the existing method of measuring the QGT that uses the pseudospins in photonic and exciton-polariton systems by suggesting a method to construct the left eigenstates from experiments. We also show that the QGT components have clear signatures in wave-packet dynamics, where the anomalous Hall drift arises from both the non-Hermitian Berry curvature and Berry connection, suggesting that both left and right eigenstates are necessary for defining non-Hermitian band geometries and topologies.

\end{abstract*}

\section{Introduction}
Topological electronic and photonic materials have attracted a lot of interest due to non-trivial surface states arising from non-zero topological invariants in the bulk of the system \cite{kane2005,kane2005graphene,xiao2010,senthil2015}. One such topological invariant is the Chern number, which measures the flux of the Berry curvature through the Brillouin zone, the unit cell of the reciprocal lattice in momentum space \cite{berry1984,chang1995,xiao2010}. The Berry curvature arises from the band structure and acts on the electrons in the solid state materials or wave packets in a photonic or hybrid photonic (polaritonic) system as an effective magnetic field, resulting in an anomalous drift with a velocity transverse to an external force or a potential gradient in real space \cite{xiao2010,chang1995,sundaram1999,chang2008,culcer2005,gao2014,gao2015WP,bleu2018WP,gianfrate2020,leblanc2021}. The Berry curvature is part of a quantity called the quantum geometric tensor (QGT)~\cite{shen2018,bleu2018,solnyshkov2021}. This quantity was introduced in Ref. \cite{provost1980} as a metric tensor that measures the change in a quantum state corresponding to infinitesimal change in the parameter space. Its real part defines the quantum metric tensor (QMT), and its imaginary part is the Berry curvature \cite{provost1980,shen2018,bleu2018,solnyshkov2021,gianfrate2020,gao2015WP,liao2021,zhang2019,zhu2021}. The quantum metric tensor is also related to the fidelity which measures the overlap between two states \cite{gu2010,jozsa1994,tzeng2021,jiang2018,sun2021,matsumoto2020,tu2023,ye2023,hetenyi2023}. 

Recently, the Berry curvature and the QMT have been generalized to dissipative quantum systems, where the gain and loss are described by non-Hermitian effective Hamiltonians \cite{el-ganainy2018,ghatak2019,bergoltz2021,ozdemir2019,shen2018,solnyshkov2021,liao2021,zhang2019,zhu2021,kunst2018}. These non-Hermitian systems exhibit unique features such as a new type of spectral degeneracy \cite{leykam2017,gao2015,krol2022,zhou2018}, as well as  novel topological invariants \cite{su2021,zhang2020} and topological edge states \cite{kunst2018,gong2018,kawabata2018,lee2019,hofmann2020,yao2018,zhang2020,zhang2022,weidemann2020}. Furthermore, although the quantum metric only appears as a non-adiabatic correction in the dynamics of Hermitian systems \cite{bleu2018WP}, it could play a dominant role in the dynamics of non-Hermitian systems \cite{solnyshkov2021}. 

Despite the increasing interest in non-Hermitian QGT and non-Hermitian Berry curvature, their definition remains ambiguous. The ambiguity arises mainly because the right and left eigenstates of the effective non-Hermitian operator are not equal. Therefore, some authors define these quantities using both left and right eigenstates \cite{shen2018,zhang2019,fan2020,zhu2021,ye2023,tzeng2021,jiang2018,sun2021,tu2023,brody2013}, while others only consider the right eigenstates \cite{shen2018,solnyshkov2021,liao2021,matsumoto2020,cuerda2023}. \red{There is no consensus on which generalization reflects reality as different works reach different conclusions. Previous studies indicate that the generalization based on both the left and right eigenstates correctly predicts phase transitions~\cite{ye2023,sun2021,zhang2019,tzeng2021,jiang2018} while the one based solely on the right eigenstates fails. However, other works indicate that the latter plays the dominant role in non-Hermitian wave-packet dynamics\cite{xu2017,silberstein2020,wang2022,solnyshkov2021}. It is therefore important to enable direct measurement of the components of the QGT to shed light on this issue.} In this work, we review these two general approaches in constructing the QGT and \red{suggest a way to experimentally measure} the QGT components. We then derive the components of the two generalized QGTs for an exemplary non-Hermitian system of microcavity exciton polaritons \cite{tercas2014,su2021}.

Exciton polaritons are composite particles arising from the strong coupling between the electron-hole pairs (excitons) and photons confined in a microcavity \cite{kasprzak2006,gao2015,deng2010,carusotto2013}. These hybrid particles are inherently non-Hermitian due to the photonic loss from the microcavity and the pump (gain) that re-injects polaritons into the system. Although non-Hermiticity is typically neglected in the studies of topological effects in exciton-polariton systems \cite{klembt2018,pieczarka2021,ren2021}, this platform has recently been used to study the novel non-Hermitian topology and observe the non-Hermitian topological invariants~\cite{su2021}. Recent works also explored non-Hermitian topological lattices~\cite{pickup2020, comaron2020}, including the role of nonlinearity~\cite{pernet2022,sigurdsson2017}, which in some configurations can lead to nonreciprocal dynamics and the non-Hermitian skin effect~\cite{mandal2020, xu2021, xu2021b, mandal2022, xu2022, kokhanchik2023}. There are also studies on the non-Hermitian eigenenergy structure, which features exceptional points, in both momentum~\cite{su2021, liao2021, krol2022}, and parameter~\cite{gao2015, gao2018, estrecho2016, li2022} spaces. \red{Furthermore, the exciton-polariton system is continuous and two-dimensional, making it an excellent platform for studying wave-packet dynamics and the effects of quantum geometry, like the anomalous Hall effect, without the complexity of the underlying periodic potential in lattices.}

It was shown theoretically that the QGT can be experimentally measured in photonic and polaritonic systems in Ref. \cite{bleu2018} and it was first experimentally observed in an exciton-polariton system in the Hermitian limit, where the gain and loss were neglected, in Ref.\cite{gianfrate2020}. \red{The method was recently applied to measure the QGT of non-Hermitian systems~\cite{liao2021,cuerda2023}. However, the formalism is based solely on right eigenstates, and hence may not apply to generic non-Hermitian systems. Therefore, it is important to experimentally access the left eigenstate to allow the construction of generalized non-Hermitian QGTs}. In this work, we consider a non-Hermitian \red{exciton-polariton model \cite{su2021,hu2022} to} extend the formalism in Ref. \cite{bleu2018} \red{and} show that \red{generalized QGTs} defined in terms of both left and right eigenstates can be experimentally measured.

It is also important to detect the direct influence of the QGT components on other observable quantities. To guide experiments in this direction, we analyze the dynamics of a wave packet in a non-Hermitian exciton-polariton system in an external field. In addition to previously reported self-acceleration of the centre of mass \cite{hu2022}, we observe the anomalous Hall drift and field-induced corrections described in previous works~\cite{xu2017,silberstein2020,wang2022}. We show that both left and right eigenstates critically contribute to this dynamics via the generalized Berry connections and Berry curvature, which highlights the importance of measuring QGT components based on both eigenstates.

This work is organized as follows. In Section \ref{sec: qgt}, we briefly revisit the QGT and Berry curvature in the conventional Hermitian systems. In Section \ref{sec: nh qgt}, we review the two different generalizations of the QGT in non-Hermitian systems. In Section \ref{sec: qgt s}, we discuss how these two generalized QGTs can be measured in a general non-Hermitian two-band model similar to the Dirac model. In Section \ref{sec: polariton}, we present our results of the generalized QGTs in a non-Hermitian exciton-polariton system. Finally, in Section \ref{sec: WP}, we analyze the wave-packet dynamics using formalism developed in previous works \cite{xu2017,silberstein2020,wang2022} and extend the formalism to investigate the interplay between the finite-size effect and the anomalous Hall drift. Our results on the generalized QGTs for a non-Hermitian Dirac model are presented in the Supplemental Document.

\section{Quantum Geometric Tensor and Berry curvature}\label{sec: qgt}
\red{The QGT and related quantities can be calculated from the system's Hamiltonian or its eigenstates. In Hermitian systems,} the QGT\cite{provost1980} arises from the distance between two quantum states \red{$|\psi_n(\zeta)\rangle$ and $|\psi_n(\zeta+d\zeta)\rangle$} and is defined as
\begin{equation}
    \begin{split}
        Q_{n,\mu\nu}=\langle\partial_\mu\psi_n|\partial_\nu\psi_n\rangle-\langle\partial_\mu\psi_n|\psi_n\rangle\langle\psi_n|\partial_\nu\psi_n\rangle
    \end{split}
\end{equation}
where $|\psi_n\rangle$ smoothly depends on the $N$-dimensional parameter $\zeta=(\zeta_1,...,\zeta_N)$, and we denote $\partial_\mu=\partial/\partial\zeta_\mu$. The first term on the right-hand side arises from expanding $|\psi_n(\zeta+d\zeta)\rangle\approx|\psi_n(\zeta)\rangle+\sum_\mu |\partial_\mu \psi_n(\zeta)\rangle d\zeta_\mu$, which results in
\begin{equation}
     ||\psi_n(\zeta+d\zeta)\rangle-|\psi_n(\zeta)\rangle||^2 \approx\langle\partial_\mu\psi_n|\partial_\nu\psi_n\rangle d\zeta_\mu d\zeta_\nu
\end{equation}
where $||\cdot||^2$ denotes the norm and the indices $\mu, \nu$ are summed over. The second term is required to guarantee that the QGT is gauge-invariant. The real part of the QGT is the quantum metric tensor (QMT) $g_{n,\mu\nu}=\operatorname{Re}{Q_{n,\mu\nu}}$ which defines the distance between two quantum states, $ds^2=g_{n,\mu\nu}d\zeta_\mu d\zeta_\nu$. Its imaginary part corresponds to the Berry curvature $\Omega_{n,\mu\nu}=-2\operatorname{Im}{Q_{n,\mu\nu}}$
\begin{equation}
    \Omega_{n,\mu\nu}=i\left(\langle\partial_{\mu}\psi_n|\partial_{\nu}\psi_n\rangle-\langle\partial_{\nu}\psi_n|\partial_{\mu}\psi_n\rangle\right)
\end{equation}
which acts as a topological effective magnetic field in momentum space~\cite{provost1980,shen2018,bleu2018,gianfrate2020,solnyshkov2021,liao2021}. In two-dimensional systems, which is the focus of this work, the only nonzero component of the Berry curvature is $\Omega_n^z=i\left(\langle\partial_{k_x}\psi_n|\partial_{k_y}\psi_n\rangle-\langle\partial_{k_y}\psi_n|\partial_{k_x}\psi_n\rangle\right)$ where its integral over the parameter space defines a topological invariant known as the Chern number~\cite{xiao2010,chang1995,berry1984,shen2018}. 

The QMT and Berry curvature can also be written in terms of the derivatives of the Hamiltonians 
\begin{equation}\label{eq: qgt pert}
    \begin{split}
        g_{n,\mu\nu}&=\operatorname{Re}\left[\sum_{m\neq n}\frac{\langle\psi_m|\partial_\mu H|\psi_n\rangle\langle\psi_n|\partial_\nu H|\psi_m\rangle}{(E_m-E_n)^2}\right]\\
        \Omega_{n,\mu\nu}&=i\sum_{m\neq n}\left[\frac{\langle\psi_m|\partial_\mu H|\psi_n\rangle\langle\psi_m|\partial_\nu H|\psi_n\rangle}{(E_m-E_n)^2}-\frac{\langle\psi_m|\partial_\nu H|\psi_n\rangle\langle\psi_n|\partial_\mu H|\psi_m\rangle}{(E_m-E_n)^2}\right]
    \end{split}
\end{equation}
which arise from perturbation theory, where $E_n$ denotes the $n$-th eigenenergy \cite{bleu2018,piechon2016}. As a result, for a two-level system, where $n=\pm$, it is easy to see that the components of the QGT follow the identities \cite{bleu2018,piechon2016}
\begin{equation}\label{eq: two level}
    \begin{split}
        g_{+,\mu\nu}&=g_{-,\mu\nu}\\
        \Omega_{+,\mu\nu}&=-\Omega_{-,\mu\nu}.
    \end{split}
\end{equation}

The QMT is also related to the fidelity $\mathcal{F}$ which measures the overlap between two quantum states, $\mathcal{F}(|\psi\rangle,|\phi\rangle)=|\langle\psi|\phi\rangle|^2$, and has been used to identify spectral degeneracies and study phase transition \cite{jozsa1994,gu2010,tzeng2021,zhang2019,tu2023,matsumoto2020,sun2021,jiang2018}. The QGT is related to the fidelity between the quantum state $|\psi_n(\zeta)\rangle$ and $|\psi_n(\zeta+d\zeta)\rangle$ expanded to second order in $d\zeta$
\begin{equation}
    \begin{split}
    \mathcal{F}(|\psi_n(\zeta)\rangle,|\psi_n(\zeta+d\zeta)\rangle)&=\langle\psi_n(\zeta)|\psi_n(\zeta+d\zeta)\rangle\langle\psi_n(\zeta+d\zeta)|\psi_n(\zeta)\rangle\\
        &\approx 1 - \chi_\mathcal{F}d\zeta_\mu d\zeta_\nu+...
    \end{split}
\end{equation}
where the leading order $\chi_\mathcal{F}$ is called the fidelity susceptibility and is an alternative expression of the QMT \cite{gu2010,ye2023,zhang2019}
\begin{equation}
    \chi_\mathcal{F}=\frac{1}{2}\left(\langle\partial_\mu\psi_n|\partial_\nu\psi_n\rangle+\langle\partial_\nu\psi_n|\partial_\mu\psi_n\rangle\right)-\langle\partial_\mu\psi_n|\psi_n\rangle\langle\psi_n|\partial_\nu\psi_n\rangle=g_{n,\mu\nu}.
\end{equation}
There are different ways \red{of defining fidelity and we summarize them in the Supplemental Document.}

Experimentally, the QGT and Berry curvature can be measured via angle \red{(or momentum)} and polarization-resolved spectroscopy \red{in photonic and hybrid photonic (exciton-polariton) systems~\cite{bleu2018}}, which provide direct optical access to the system's eigenenergies and eigenstates. For such measurements, it is important to use systems that have polarization-dependent energy bands. In exciton-polariton systems, in particular, photons are strongly coupled to an exciton, and the polarization of light emitted as the polaritons decay is used to probe their pseudospin characterised by the Stokes vector components: $S_x=(I_H-I_V)/(I_H+I_V)$, $S_y=(I_D-I_A)/(I_D+I_A)$ and \red{$S_z=(I_\circlearrowleft-I_\circlearrowright)/(I_\circlearrowleft+I_\circlearrowright)$, where $I_{H,V,D,A}$ are the intensities of the horizontally, vertically, diagonally, and antidiagonally  polarized emissions, respectively, and $I_{\circlearrowleft,\circlearrowright}$ }are the intensities for the two circular polarizations \cite{bleu2018}.

The Stokes vector $\mathbf{S}_n(\mathbf{k})$ of the emitted photon has the same orientation as the pseudospin vector of the polariton in the Bloch sphere so we use them interchangeably here \cite{bleu2018}.  From the momentum-dependent pseudospin $\mathbf{S}_n(\mathbf{k})$ of the $n$-th \red{band}, the QGT can then be calculated as described in Ref. \cite{bleu2018} and summarized below.

The Stokes parameters can be used to reconstruct the spinor wave function or the orientation of the pseudospin in the Bloch sphere
\begin{equation}
    \begin{split}
        |\psi\rangle=\begin{pmatrix}e^{-i\phi}\cos\frac{\theta}{2}\\ \sin\frac{\theta}{2}\end{pmatrix}
    \end{split}
\end{equation}
where $\theta=\arccos S_z$, $\phi=\arctan(S_y/S_x)$ denote the angles of the pseudospin in the Bloch sphere. Following Ref.~\cite{bleu2018} the \red{QMT} and the Berry curvature can be calculated as:
\begin{equation}
\label{eq: QGT_H}
    \begin{split}
    g_{n,\mu\nu}=&\frac{1}{4}\Big(\partial_\mu\theta\partial_\nu\theta+\sin^2\theta\partial_\mu\phi\partial_\nu\phi\Big),\\
    \Omega_{n,\mu\nu}=&\frac{1}{2}\sin\theta\Big(\partial_{\mu}\phi\partial_{\nu}\theta-\partial_{\nu}\phi\partial_{\mu}\theta\Big).
    \end{split}
\end{equation}
These formulas provide a practical method for measuring these two quantities~\cite{bleu2018}, which has been demonstrated in several experiments using different photonic and polaritonic systems~\cite{gianfrate2020, ren2021, liao2021, polimeno2021tuning, lempicka2022electrically}.

However, in non-Hermitian systems, the eigenstates are generally not orthogonal, and the left and right eigenstates are not equal. Hence, there is more than one way of defining the QGT and related quantities, including the Berry curvature and the fidelity \cite{shen2018,fan2020,liao2021,zhang2019,zhu2021,cuerda2023,ye2023,matsumoto2020,sun2021,tzeng2021,tu2023}. In this work, we consider two different non-Hermitian generalizations of both the QGT and Berry curvature and suggest a protocol for extracting both of them from the measurable polarization texture of the eigenstates in an exciton-polariton system.

\section{Non-Hermitian Quantum Geometric Tensors}\label{sec: nh qgt}
In non-Hermitian systems, four different combinations of the left and right eigenstates, namely left-right (LR), right-left (RL), left-left (LL) and right-right (RR), give rise to four different ways to define the Berry curvature
\begin{equation}
\Omega_{n,\mu\nu}^{pq}=i\Big(\langle\partial_\mu\psi_n^p|\partial_\nu\psi_n^q\rangle-\langle\partial_\nu\psi_n^p|\partial_\mu\psi_n^q\rangle\Big)
\end{equation}
where $p,q=L,R$ denote the left and right eigenstates, which are normalized as $\langle\psi_n^p|\psi_n^q\rangle=1$~\cite{shen2018}. \red{Note that} these four Berry curvatures are locally different but their integrals all yield the same Chern number \cite{shen2018}.

Similar to the generalized Berry curvatures, there are also different approaches to generalize the quantum metric and the fidelity to non-Hermitian systems. As we have previously seen, the definition of QGT comes from the distance between quantum states \cite{provost1980}. In a normed vector space, the distance between two vectors is defined using a norm $||\cdot||$, defined as the inner product $\langle\cdot\rangle$ of the quantum states in quantum mechanics. Therefore, the two different definitions arise from different ways to define the inner product in non-Hermitian quantum mechanics.

In the first approach, the inner product is assumed to be the same as that in the Hermitian systems. \red{To avoid confusion with notations, we denote states normalized this way with a bar such that $\langle\bar{\psi}^p|\bar{\psi}^p\rangle=1$}. In this case, the distance between the states $|\bar{\psi}_n^R(\lambda+d\lambda)\rangle$ and $|\bar{\psi}_n^R(\lambda)\rangle$ yields a similar results to the Hermitian case, i.e.
$$||\bar{\psi}_n^R(\zeta+d\zeta)\rangle-|\bar{\psi}_n^R(\zeta)\rangle||^2\approx||\partial_\mu\bar{\psi}_n^R(\zeta)\rangle d\zeta_\mu||^2=\langle\partial_\mu\bar{\psi}_n^R|\partial_\nu\bar{\psi}_n^R\rangle d\zeta_\mu d\zeta_\nu.$$
After including terms that restore gauge invariance \cite{provost1980}, this gives rise to the \textit{right-right (RR) QGT}
\begin{equation}\label{eq: RR QGT}
    Q_{n,\mu\nu}^{RR}=\langle\partial_\mu\bar{\psi}_n^R|\partial_\nu\bar{\psi}_n^R\rangle-\langle\partial_\mu\bar{\psi}_n^R|\bar{\psi}_n^R\rangle\langle\bar{\psi}_n^R|\partial_\nu\bar{\psi}_n^R\rangle.
\end{equation}
This \red{RR} QGT is a Hermitian tensor and its imaginary part corresponds to the RR Berry curvature $\Omega_{n,\mu\nu}^{RR}=-2\operatorname{Im}{Q_{n,\mu\nu}^{RR}}$. Therefore, one can intuitively define the RR QMT as its real part $g_{n,\mu\nu}^{RR}=\operatorname{Re}{Q_{n,\mu\nu}^{RR}}$ \cite{solnyshkov2021,liao2021}.

The second approach defines the inner product using both the left and right eigenstates. Some works use the biorthogonal approach, where the inner product of a state $|\psi\rangle$ with itself is defined as $\langle\hat{\psi}|\psi\rangle$, with $\langle\hat{\psi}|$ denoting associate states \cite{brody2013,brody2014}
\begin{equation}
    \begin{split}
        |\psi\rangle&=\sum_n c_n|\psi_n^R\rangle\\
        \langle\hat{\psi}|&=\sum_n c_n^*\langle\psi_n^L|.
    \end{split}
\end{equation}
Other works use the metricized approach, where similar results are obtained by defining the inner product through a metric operator which maps $\langle\psi_n^R|$  to $\langle\psi^L_n|$ \cite{tzeng2021,zhang2019,gardas2016,mostafazadeh2007,ju2019,tu2023}.
Using this definition of inner product in the biorthogonal approach or the metricized approach, the distance between the states $|\psi_n^R(\zeta+d\zeta)\rangle$ and $|\psi_n^R(\zeta)\rangle$ is now
$$||\psi_n^R(\zeta+d\zeta)\rangle-|\psi_n^R(\zeta)\rangle||^2\approx||\partial_\mu\psi_n^R(\zeta)\rangle d\zeta_\mu||^2=\langle\partial_\mu\psi_n^L(\zeta)|\partial_\nu\psi_n^R(\zeta)\rangle d\zeta_\mu d\zeta_\nu.$$
After restoring terms preserving gauge invariance, this yields the \textit{left-right (LR)} QGT
\begin{equation}\label{eq: LR QGT}
Q_{n,\mu\nu}^{LR}=\langle\partial_\mu\psi_n^L|\partial_\nu\psi_n^R\rangle-\langle\partial_\mu\psi_n^L|\psi_n^R\rangle\langle\psi_n^L|\partial_\nu\psi_n^R\rangle.
\end{equation}
which is \textit{no longer Hermitian}~\cite{brody2013,zhang2019,zhu2021,ye2023}. The anti-symmetric part of the LR QGT corresponds to the LR Berry curvature $\Omega_{n,\mu\nu}^{LR}=i(Q_{n,\mu\nu}^{LR}-Q_{n,\nu\mu}^{LR})$, which is now a complex-valued quantity \cite{shen2018,fan2020}. 

In $PT$-symmetric \cite{zhang2019} or pseudo-Hermitian \cite{zhu2021} systems, the LR QMT can be defined as the real and symmetric part of the LR QGT ${g}_{n,\mu,\nu}^{LR}=\operatorname{Re}\left[\frac{1}{2}(Q_{n,\mu\nu}^{LR}+Q_{n,\mu\nu}^{RL})\right]$. However, it is still unclear which part of the LR QGT corresponds to the LR QMT in a general non-Hermitian system. In Ref.~\cite{ye2023}, the authors proposed three different generalizations, namely
\begin{equation}\label{eq: 3 lr qmt}
    \begin{split}
        g_{n,\mu\nu}^{LR}&=\frac{1}{2}(Q_{n,\mu\nu}^{LR}+Q_{n,\nu\mu}^{LR}),\quad \textnormal{(symmetric part)}\\
        g_{n,\mu\nu}^{LR}&=\operatorname{Re}\left[Q_{n,\mu\nu}^{LR}\right],\quad \textnormal{(real part)}\\
        g_{n,\mu\nu}^{LR}&=\operatorname{Re}\left[\frac{1}{2}(Q_{n,\mu\nu}^{LR}+Q_{n,\mu\nu}^{RL})\right].\quad \textnormal{(real and symmetric part)}
    \end{split}
\end{equation}

Similar to the QGT, there are also different ways to generalize the fidelity and fidelity susceptibility to non-Hermitian systems~\cite{matsumoto2020,sun2021,tzeng2021,zhang2019,jiang2018,tu2023}. \red{See the Supplemental Document for a comparison of the different generalizations. Since fidelity is not the focus of this work, we refer to all of them simply as LR fidelity.}

Despite the vast interest in the QGT and fidelity in non-Hermitian systems, there is no consensus on which one is the correct generalization, and different works reach different conclusions. \red{On one hand, the authors in Refs.~\cite{xu2017,silberstein2020,wang2022}} showed that the anomalous Hall drift of the wave packets in non-Hermitian systems can be described using the RR Berry curvature. In Ref.~\cite{solnyshkov2021}, the authors also suggested that the RR quantum metric plays an important role in the wave-packet dynamics in non-Hermitian systems.

On the other hand, other works suggested that the LR quantum metric and fidelity susceptibility correctly predict phase transitions \cite{ye2023,sun2021,zhang2019,tzeng2021,jiang2018} while the RR ones fail \cite{ye2023,sun2021}. This is because the divergences of the fidelity and fidelity susceptibility signal the presence of spectral degeneracies, which are required for a phase transition to occur \cite{jozsa1994,gu2010,tzeng2021,zhang2019,tu2023,matsumoto2020,sun2021,jiang2018}. However, contrary to these claims, in Ref.~\cite{matsumoto2020}, the authors observed a divergent RR fidelity susceptibility in the absence of spectral degeneracy, this effect signifying that non-Hermitian systems can undergo phase transition without gap closing.

\red{Experiments are therefore needed to shed more light on the above discrepancies and to establish the relevance of different generalizations to the experimentally observable effects. To date, only} the RR QMT \cite{liao2021} and Berry curvature \cite{cuerda2023} have been experimentally measured by using the formalism in Ref.~\cite{bleu2018}. However, this method only enables the measurements of the RR QGT. In the following, we propose a protocol for experimental measurement of the LR QGT in a two-band system, e.g., a photonic or exciton-polariton system with polarization-dependent energy splitting.

For simplicity, we refer to the first definition in Eq.~(\ref{eq: 3 lr qmt}), i.e., the symmetric part of the LR QGT, as the LR QMT, since one can easily reconstruct the other two definitions from it. The real and symmetric part of the LR QGT is simply $\operatorname{Re}\left[g_{n,\mu\nu}^{LR}\right]$ while the real part is
$\operatorname{Re}\left[g_{n,\mu\nu}^{LR}\right]+\frac{1}{2}\operatorname{Im}\left[\Omega_{n,\mu\nu}^{LR}\right]$. Moreover, when $g_{n,\mu\nu}^{LR}$ is defined as the symmetric part of the LR QGT, its  structure and properties are similar to those of the Hermitian QMT, which we will later demonstrate for an exciton-polariton system.

\section{Measuring Non-Hermitian Quantum Geometric Tensor}\label{sec: qgt s}
We consider a general two-band non-Hermitian Hamiltonian in momentum ($k$) space:
\begin{equation}\label{eq: general h}
    \begin{split}
        \mathbf{H}(\mathbf{k})&=h_0(\mathbf{k}) \mathbb{I}_{2\times 2} + \mathbf{h}(\mathbf{k})\cdot{\boldsymbol\sigma}\\
        \mathbf{h}(\mathbf{k})&=[h_x(\mathbf{k}),h_y(\mathbf{k}),h_z(\mathbf{k})]
    \end{split}
\end{equation}
where $\mathbb{I}_{2\times 2}$ is the 2$\times$2 identity matrix, $\boldsymbol{\sigma}$ is a vector of Pauli matrices $\sigma_j$ representing pseudospins in photonic or polaritonic systems. Here, ${\mathbf{h}}$ is a complex-valued effective magnetic field acting on the pseudospin. The $h_0(\mathbf{k})$ component captures the energy offset, such as the kinetic energy, which does not affect the QGT.

The complex-valued eigenenergies of these models take the general form $E_\pm = h_0 \pm \lambda$, where $\lambda=\sqrt{h_x^2+h_y^2+h_z^2}$. In the following, we drop the $\mathbf{k}$-dependence for brevity. The corresponding un-normalized left and right eigenstates of the Hamiltonian in Eq. (\ref{eq: general h}) are
\begin{equation}\label{eq: eigenstates_LR}
    |R_\pm\rangle=\begin{pmatrix}h_z\pm \lambda\\h_+\end{pmatrix}
    \qquad \textnormal{and} \qquad
    |L_\pm\rangle=\begin{pmatrix}h_z^*\pm \lambda^*\\h_-^*\end{pmatrix}.
\end{equation}
where $h_\pm=h_x\pm i h_y$. These two sets of eigenstates result in at least two generalizations of the QGT and their components. In the following, we show how $Q^{RR}_{\pm,\mu\nu}$ and $Q^{LR}_{\pm,\mu\nu}$ can be constructed from experimentally accessible observables.

\subsection{Right-right Quantum Geometric Tensor}
The pseudospins components of a photonic or exciton-polariton system can be measured from the polarization of light, i.e. from the Stokes vector. In exciton-polariton systems, we only have access to the polarization of the right eigenstates which corresponds to the Stokes vectors as follows:
\begin{equation}\label{eq: S_RR}
    \begin{split}
S_{j,\pm}^{R}&=\frac{\langle R_\pm|\sigma_j|R_\pm\rangle}{\langle R_\pm|R_\pm\rangle}
    \end{split}
\end{equation}
where $j=\{x,y,z\}$. 
As discussed in Section 2, this definition coincides with how pseudospins are usually measured in exciton-polariton systems using the intensities of different polarizations~\cite{bleu2018}. It also gives real-valued pseudospins which have been experimentally measured  \cite{gianfrate2020,su2021,liao2021}. Moreover, it matches the non-antipodal behaviour of the pseudospins on the Bloch sphere, which reflects the non-orthogonality of the eigenstates~\cite{su2021}.

From these experimentally measured components, we can define the angles on the Bloch sphere as $\theta_\pm^{R}=\arccos S_{z,\pm}^{R}$, $\phi_\pm^{R}=\arctan(S_{y,\pm}^{R}/ S_{x,\pm}^{R})$. Then, using the formalism presented in Ref. \cite{bleu2018}, the spinor wavefunction defined using these two angles recovers the right eigenstates and their conjugates
\begin{equation}\label{eq: R psi}
\begin{split}
    \begin{pmatrix}e^{-i\phi_\pm^{R}}\cos\frac{\theta_\pm^{R}}{2}\\ \sin\frac{\theta_\pm^{R}}{2}\end{pmatrix}
    &=\frac{1}{\sqrt{\langle R_\pm|R_\pm\rangle}}\sqrt{\frac{h_+^*}{h_+}}|R_\pm\rangle=|\bar{\psi}^R\rangle\\
    \begin{pmatrix}e^{i\phi_\pm^{R}}\cos\frac{\theta_\pm^{R}}{2} & \sin\frac{\theta_\pm^{R}}{2}\end{pmatrix}
    &=\frac{1}{\sqrt{\langle R_\pm|R_\pm\rangle}}\sqrt{\frac{h_+}{h_+^*}}\langle R_\pm|=\langle\bar{\psi}^R_\pm|.
\end{split}
\end{equation}
\red{In this approach, the right eigenstates $|\bar{\psi}^R_\pm\rangle$ are normalized as} $\langle\bar{\psi}^R_\pm|\bar{\psi}^R_\pm\rangle=1$. Plugging the left-hand side back to the definition of the RR QGT in Eq.~(\ref{eq: RR QGT}), we can write down the components of the RR QGT as:
\begin{equation}\label{eq: QRR S}
    \begin{split}
        g^{RR}_{\pm,\mu\nu}&=\frac{1}{4}\Big(\partial_\mu\theta_\pm^{R}\partial_\nu\theta_\pm^{R}+\sin^2\theta_\pm^{R} \partial_\mu\phi_\pm^{R}\partial_\nu\phi_\pm^{R}\Big)\\
        \Omega^{z,RR}_{\pm,z}&=\frac{1}{2}\sin\theta_\pm^{R}\Big(\partial_{k_x}\phi^{R}_\pm\partial_{k_y}\theta^{R}_\pm-\partial_{k_y}\phi^{R}_\pm\partial_{k_x}\theta^{R}_\pm\Big).
    \end{split}
\end{equation}
This is the same result as the Hermitian case Eqs.~(\ref{eq: QGT_H}) showing that the formalism from Ref.~\cite{bleu2018} can be directly applied to non-Hermitian systems \red{yielding} the RR QMT and Berry curvature~\cite{liao2021, cuerda2023}.

\subsection{Left-Right Quantum Geometric Tensor}
\begin{figure}[h]
    \centering
    \includegraphics[width=0.95\textwidth]{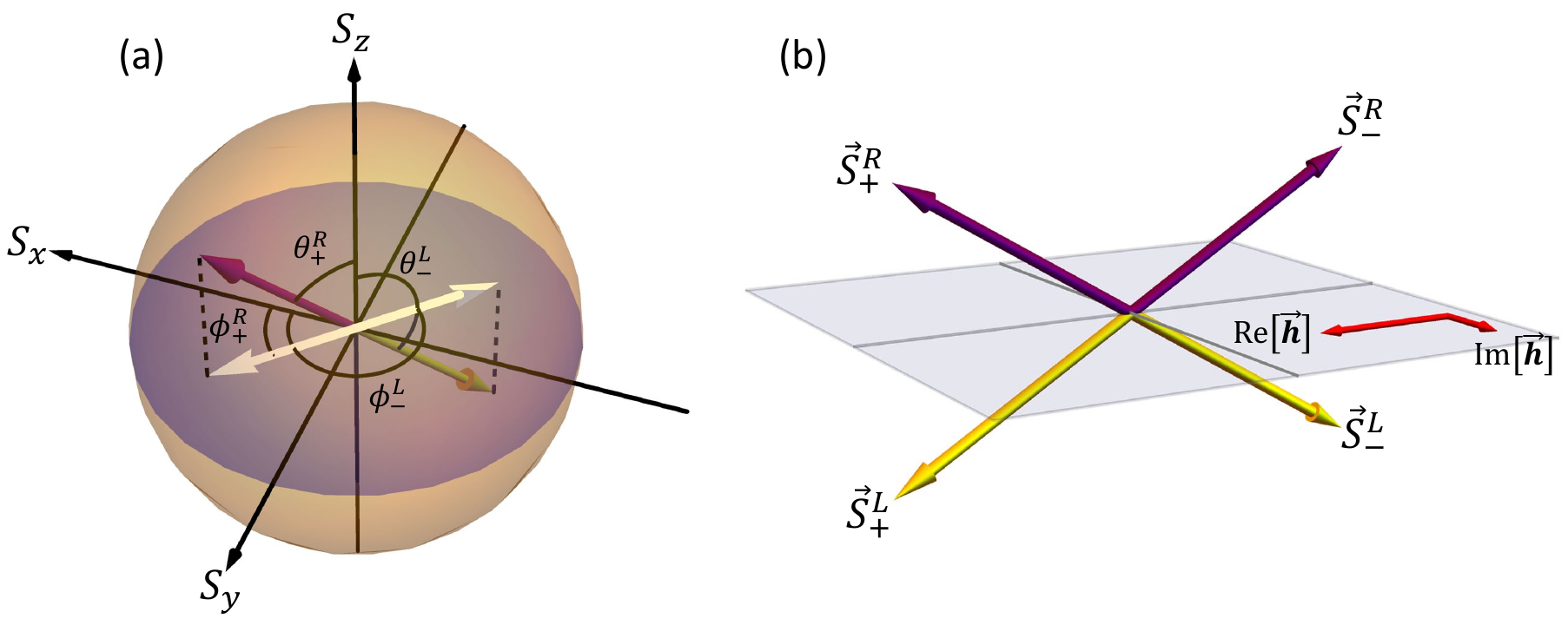}
    \caption{The pseudospin of the right (purple) and left (yellow) eigenstates with their projections (white) on the $[S_x,S_y]$-plane plotted with (a) the Bloch sphere. (b) Same as (a) but showing all four pseudospins oriented with respect to the real and imaginary parts of the effective magnetic field (red). The pseudospin $\mathbf{S}_\pm^R$ points in the opposite direction to $\mathbf{S}^L_\mp$, while the real and imaginary parts of the effective magnetic field define the \red{mirror symmetry} plane.}
    \label{fig: S}
\end{figure}

Measuring the LR QGT is not as straightforward since the pseudospins from both the left and the right eigenstates are required. However, \red{at least for the system considered here, experiments} only have direct access to the right eigenstates. Fortunately, in a two-band system, the left eigenstates are closely related to the right eigenstate by the bi-orthonormality condition $\langle\psi_\pm^L|\psi_\pm^R\rangle=1$ and $\langle\psi_\pm^L|\psi_\mp^R\rangle=0$. \red{The biorthogonality means that} the left and right eigenstates with different indices are antipodal on the Bloch sphere, \red{as shown in Fig.~\ref{fig: S}. The right and left eigenstates are actually mirrored images of each other where} the plane of symmetry is formed by the real and imaginary parts of the effective field $\mathbf{h}$. Hence, the relation between the eigenstate pseudospins are simply:
\begin{equation}
   \mathbf{S}_{\pm}^L=-\mathbf{S}_{\mp}^R,
\end{equation}
where the components of $\mathbf{S}_{\pm}^L$ are defined as $S_{j,\pm}^L=\langle L_\pm^L|\sigma_j| L_\pm^L\rangle/\langle L_\pm^L| L_\pm^L\rangle$.

The angles of the pseudospins of the left eigenstates in the Bloch sphere can then be written in terms of the right eigenstates as $\phi_\pm^L=\pi+\phi^R_\mp$, $\theta_\pm^L=\pi-\theta^R_\mp$. Hence, the left eigenstates can be reconstructed using the pseudospin textures \red{(or the Bloch sphere angles)} of the right eigenstates as:
\begin{equation}\label{eq: L psi}
    \begin{split}
       \begin{pmatrix} 
            -e^{i\phi_\mp^R}\sin\frac{\theta_\mp^R}{2} & \cos\frac{\theta_\mp^R}{2}
        \end{pmatrix}
        =\frac{1}{\sqrt{\langle L_\pm|L_\pm\rangle}}\sqrt{\frac{h_-^*}{h_-}}\langle L_\pm|=\langle\bar{\psi}^L_\pm|
    \end{split}
\end{equation}
where $\langle\bar{\psi}^L_\pm|\bar{\psi}^L_\pm\rangle=1$.

Given both the right and left eigenstates, we can then construct the spin components of the biorthogonal $\mathbf{S}_\pm^{LR}$ which we use to \red{construct the LR} QGT, $Q_{\pm,\mu\nu}^{LR}$. Similarly to the RR version [see Eqs.~(\ref{eq: S_RR})], the pseudospin components are
\begin{equation}\label{eq: BO S}
    \begin{split}
S_{j,\pm}^{LR}&=\frac{\langle\bar{\psi}^{L}_\pm|\sigma_j|\bar{\psi}^{R}_\pm\rangle}{\langle\bar{\psi}^{L}_\pm|\bar{\psi}^{R}_\pm\rangle}
    \end{split}.
\end{equation}
where $j=\{x,y,z\}$, following the biorthogonal formalism in Refs. \cite{brody2014,brody2013,gardas2016,ju2019,mostafazadeh2007,liu2019}. Note that the components of $\mathbf{S}_\pm^{LR}$ are complex-valued, but they are constructed here from the real-valued components of $\mathbf{S}_\pm^{R}$ using Eqs.~(\ref{eq: L psi}) and~(\ref{eq: BO S}). \red{The general forms of the $\mathbf{S}^R$  and $\mathbf{S}^{LR}$ for any two-band non-Hermitian Hamiltonian, Eq.~(\ref{eq: general h}), are presented in the Supplemental Document.}

Explicitly, the components of $\mathbf{S}^{LR}_\pm$ can be written in terms of the measured real-valued pseudospin angles $\phi_\pm^R$ and $\theta_\pm^R$ as
\begin{equation}\label{eq: SLR_from_R}
    \begin{split}
        S_{x,\pm}^{LR}&=\frac{e^{-i\phi_\pm^R}\cos{\frac{\theta_\mp^R}{2}}\cos{\frac{\theta_\pm^R}{2}}-e^{i\phi_\mp^R}\sin{\frac{\theta_\mp^R}{2}}\sin{\frac{\theta_\pm^R}{2}}}{S_{0,\pm}}\\
        S_{y,\pm}^{LR}&=\frac{i e^{-i\phi_\pm^R}\cos{\frac{\theta_\mp^R}{2}}\cos{\frac{\theta_\pm^R}{2}}+i e^{i\phi_\mp^R}\sin{\frac{\theta_\mp^R}{2}}\sin{\frac{\theta_\pm^R}{2}}}{S_{0,\pm}}\\
        S_{z,\pm}^{LR}&=\frac{-e^{i(\phi_\mp^R-\phi_\pm^R)}\cos{\frac{\theta_\pm^R}{2}}\sin{\frac{\theta_\mp^R}{2}}-\cos{\frac{\theta_\mp^R}{2}}\sin{\frac{\theta_\pm^R}{2}}}{S_{0,\pm}}\\
        S_{0,\pm} &= -e^{i(\phi_\mp^R-\phi_\pm^R)}\cos{\frac{\theta_\pm^R}{2}}\sin{\frac{\theta_\mp^R}{2}}+\cos{\frac{\theta_\mp^R}{2}}\sin{\frac{\theta_\pm^R}{2}} = \langle\bar\psi_\pm^L |\bar\psi_\pm^R \rangle
    \end{split}
\end{equation}
From these complex-valued pseudospin components, we can calculate the corresponding complex angles on the Bloch sphere \cite{bleu2018} , $\theta_\pm^{LR}=\arccos S_{z,\pm}^{LR}$ and $\phi_\pm^{LR}=\arctan(S_{y,\pm}^{LR}/ S_{x,\pm}^{LR})$, which are:
\begin{equation}
    \begin{split}
        \phi_\pm^{LR}&=-i\operatorname{tanh}^{-1}\Bigg(1+\frac{2}{e^{i(\phi_\mp^R+\phi_\pm^R)}\tan{\frac{\theta_\mp^R}{2}}\tan{\frac{\theta_\pm^R}{2}}-1}\Bigg)\\
        \theta_\pm^{LR}&=\cos^{-1}{\Bigg(\frac{2}{1-e^{i(\phi_\pm^R-\phi_\mp^R)}\cot{\frac{\theta_\mp^R}{2}}\tan{\frac{\theta_\pm^R}{2}}}-1\Bigg)}.
    \end{split}
\end{equation}

Using these $\theta_\pm^{LR}$, $\phi_\pm^{LR}$, the bi-orthonormal left and right eigenstates are reconstructed as:
\begin{equation}
\begin{split}
    \begin{pmatrix}e^{-i\phi_\pm^{LR}}\cos\frac{\theta_\pm^{LR}}{2}\\ \sin\frac{\theta_\pm^{LR}}{2}\end{pmatrix}
    &=\frac{1}{\sqrt{\langle L_\pm|R_\pm\rangle}}\sqrt{\frac{h_-}{h_+}}|R_\pm\rangle=|\psi^R_\pm\rangle\\
    \begin{pmatrix}e^{i\phi_\pm^{LR}}\cos\frac{\theta_\pm^{LR}}{2} & \sin\frac{\theta_\pm^{LR}}{2}\end{pmatrix}
    &=\frac{1}{\sqrt{\langle L_\pm|R_\pm\rangle}}\sqrt{\frac{h_+}{h_-}}\langle L_\pm|=\langle\psi^L_\pm|.
\end{split}
\end{equation}
These can also be written in terms of the original angles, $\theta^R_\pm$ and $\phi^R_\pm$ , by normalizing Eqs. ~(\ref{eq: R psi}) and (\ref{eq: L psi}) with respect to $\sqrt{\langle \bar{\psi}^L_\pm|\bar{\psi}^R_\pm\rangle}$, whose form is presented as $S_{0,\pm}^LR$ in Eqs.~(\ref{eq: SLR_from_R}).

We emphasize that the denominators in Eq.~(\ref{eq: BO S}) are necessary to recover the bi-orthonormality condition of the left and right eigenstates $\langle\psi^L_\pm|\psi^R_\pm\rangle=1$, which are required for the LR QGT. Plugging the left-hand-side of $|\psi^R_\pm\rangle$, $\langle\psi^L_\pm|$ in the LR QGT definition in Eq.~(\ref{eq: LR QGT}), the components of the LR QGT can be expressed as:
\begin{equation}\label{eq: nh qgt}
    \begin{split}
        g^{LR}_{\pm,\mu\nu}=&\frac{1}{4}\left(\partial_\mu\theta_\pm^{LR}\partial_\nu\theta_\pm^{LR}+\sin^2\theta_\pm^{LR} \partial_\mu\phi_\pm^{LR}\partial_\nu\phi_\pm^{LR}\right)\\
        \Omega^{z,LR}_\pm=&\frac{1}{2}\sin\theta_\pm^{LR}\left(\partial_{k_x}\phi^{LR}_\pm\partial_{k_y}\theta^{LR}_\pm-\partial_{k_y}\phi^{LR}_\pm\partial_{k_x}\theta^{LR}_\pm\right).
    \end{split}
\end{equation}
This form is similar to the Hermitian case Eqs.~(\ref{eq: QGT_H}) and the RR QGT given by \red{Eq.~(\ref{eq: RR QGT})}, but uses the complex angles $\theta_\pm^{LR}$ and $\phi_\pm^{LR}$. Interestingly, the LR QGT can also be written explicitly in terms of the Bloch angles of the right eigenstates as
\begin{equation}\label{eq: nh qg2t}
Q^{LR}_{\pm,\mu\nu}=\frac{e^{i(\phi_\mp^R+\phi_\pm^R)}\left(\partial_\mu\theta_\mp^{R}+i\sin\theta_\mp^R\partial_\mu\phi_\mp^R\right) \left(\partial_\nu\theta_\pm^{R}+i\sin\theta_\pm^R\partial_\nu\phi_\pm^R\right)}{4\left[e^{i\phi_\mp^R} \cos(\theta_\pm^R/2) \sin(\theta_\mp^R/2) - e^{i\phi_\pm^R} \cos(\theta_\mp^R/2) \sin(\theta_\pm^R/2)   \right]^2}.
\end{equation}
Unfortunately, it is hard to extract a simple expression of the LR QMT and Berry curvature from this form.

The components of \red{$\mathbf{S}_{j,\pm}^{LR}$ and the angles} $\theta_\pm^{LR}$, $\phi_\pm^{LR}$ are generally complex-valued. However, since they can be calculated from the experimentally measurable polariton pseudospins, Eq.~(\ref{eq: nh qgt}) provides a way to experimentally determine the components of $Q_{\pm,\mu\nu}^{LR}$.

In exciton-polariton systems, only the components of $\mathbf{S}^R_\pm$ \cite{su2021} and $Q^{RR}_{\pm,\mu\nu}$ \cite{liao2021} have been experimentally measured so far, and the left eigenstates are not directly accessible. Here, we present a formalism that allows us to reconstruct the components of $\mathbf{S}_\pm^{LR}$ and $Q^{LR}_{\pm,\mu\nu}$, by measuring the pseudospins of the right eigenstates. In the following section, we will demonstrate both generalizations of non-Hermitian Berry curvature and quantum metric tensor using a non-Hermitian exciton-polariton model.

\section{Non-Hermitian Quantum Geometric Tensor in an Exciton-Polariton System}\label{sec: polariton}
To demonstrate how the \red{non-Hermitian QGTs can be measured} in a realistic system, we consider a low-energy effective model describing the lower polariton branch with polarization splitting arising from cavity anisotropy, the photonic spin-orbit coupling and the Zeeman splitting induced by an external magnetic field~\cite{su2021, hu2022}. 

\begin{figure}[h!]
    \centering
    \includegraphics[width=0.7\textwidth]{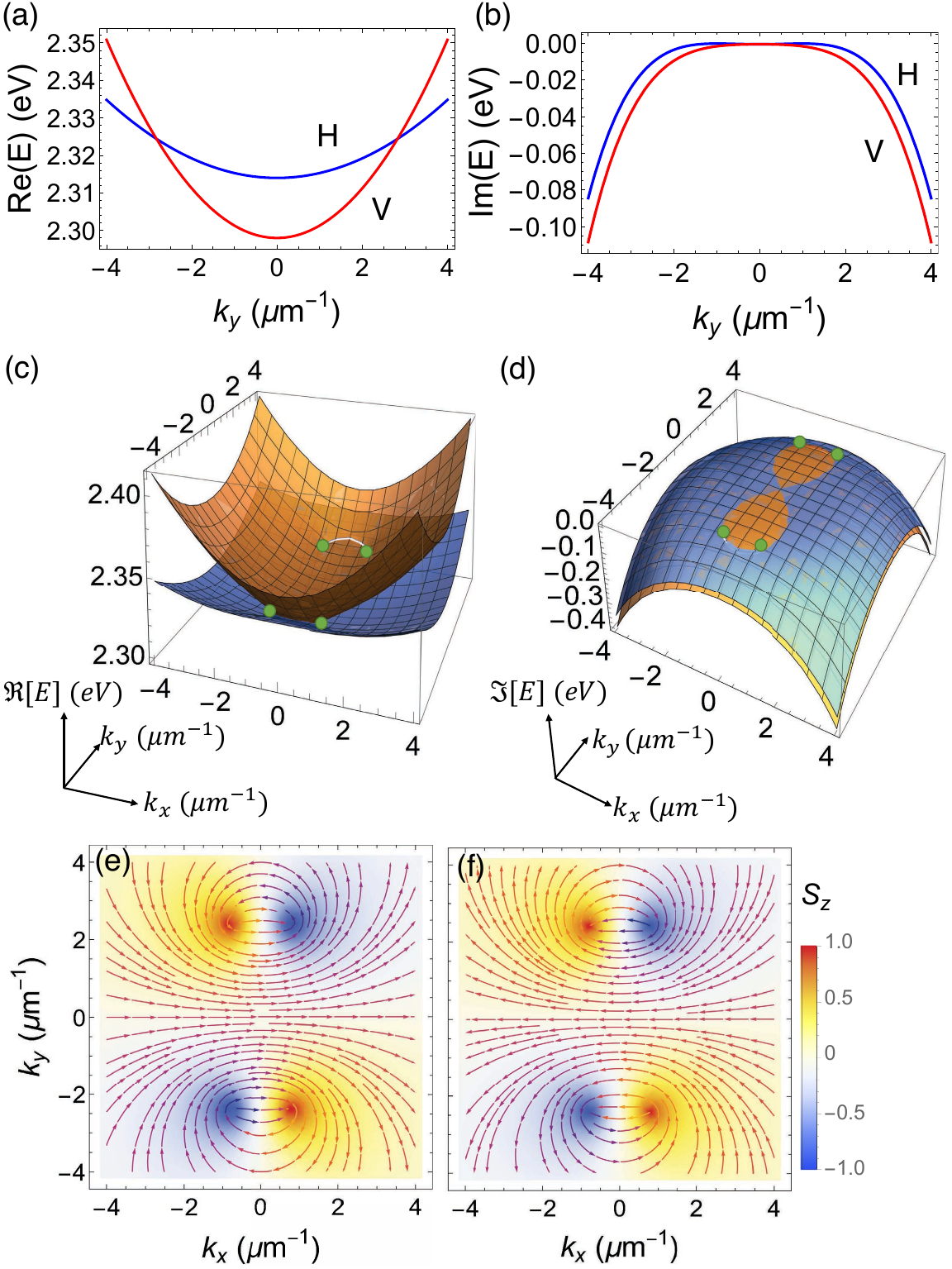}
    \caption{(a) and (b) show the splitting in the energies and linewidths of the horizontally (H) and vertically (V) polarized modes which is used to define the pseudospins. Also showing the real (c) and imaginary (d) parts of the exciton-polariton bands. The green dots denote the exceptional points. The last two panels show the pseudospin textures of the upper (e) and the lower (f) right eigenstates at $\Delta=0$, where the color denotes the out-of-plane components $S_z$ and the arrows denote the in-plane components $(S_x,S_y)$. }
    \label{fig: extra}
\end{figure}

The non-Hermitian exciton-polariton Hamiltonian $\mathbf{H}_{pol}(\mathbf{k})=h_0(\mathbf{k})\mathbb{I}_{2\times2}+\mathbf{h}(\mathbf{k})\cdot\boldsymbol\sigma$ takes the form:
\begin{equation}\label{eq: polariton H}
    \begin{split}
        h_0(\mathbf{k})&=\left(E_0+\frac{\hbar^2k^2}{2m}-i\gamma(\mathbf{k})\right)\\
        \mathbf{h}(\mathbf{k})&=[\tilde\alpha+\tilde\beta(k_x^2-k_y^2),2\tilde\beta k_xk_y,\Delta].
    \end{split}
\end{equation}
Here, $E_0$ denotes the mean energy, $m$ denotes the effective exciton-polariton mass, and $\gamma(\mathbf{k})$ denotes the \red{mean momentum-dependent linewidth. The complex parameters $\tilde\alpha=\alpha-ia$ and $\tilde\beta=\beta-ib$  account for the optical anisotropy of the birefringent microcavity that splits linearly polarized modes and the photonic spin-orbit coupling which splits the TE and TM modes, respectively. The real part directly affects the energy splitting while the imaginary part affects the difference in linewidths. In addition, $\Delta$ is a real parameter} representing the Zeeman splitting induced by an out-of-plane magnetic field. In what follows, we assume the physical values of the parameters of this system consistent with the ones presented in the Supplemental Document unless specified otherwise.

\red{The two complex exciton-polariton band eigenenergies are $E_\pm = E_0(\mathbf{k})\pm \lambda(\mathbf{k})$, where 
\begin{equation}
        \lambda(\mathbf{k})=\sqrt{\tilde\alpha^2+2\tilde\alpha\tilde\beta(k_x^2-k_y^2)+\tilde\beta^2 k^4+\Delta^2}.
\end{equation}
}The corresponding left and right eigenstates are given by Eqs.~(\ref{eq: eigenstates_LR}), with 
$h_\pm=\tilde\alpha+\tilde\beta(k_x\pm i k_y)^2$ and $h_z=\Delta$. The momentum dependence of real and imaginary parts of the bands along $k_x=0$ is shown in Figs.~\ref{fig: extra}(a,b) for the  $\Delta=0$ case. Figures~\ref{fig: extra}(c,b) show the band energies in the full 2D momentum space, where four non-Hermitian degeneracies (exceptional points) can be seen~\cite{su2021, hu2022}  at $k^{EP}=\sqrt{|\tilde\alpha/\tilde\beta|}$, $\varphi_k^{EP}=\pm\pi/2\pm\arg\sqrt{\tilde\alpha/\tilde\beta}$, in polar coordinates. When a Zeeman splitting $\Delta$ is present, it pulls the exceptional points in a pair towards each other, and at sufficiently strong Zeeman splitting, $|\Delta_c|=|(b\alpha-a\beta)/\beta|$, the exceptional points pairs merge into hybrid points \cite{shen2018}. Further increasing $\Delta$ will annihilate the degeneracies and open a gap in the exciton-polariton bands~\cite{su2021, hu2022}. 

\red{We can calculate the pseudospins from the Stokes vector (or polarization) of the eigenstates using Eq.~(\ref{eq: S_RR}). For example, the bands shown in Figs.~\ref{fig: extra}(a,b) are H and V polarized, which correspond to pseudospins oriented at antipodal directions on the equator of the Bloch sphere. The full 2D momentum-space pseudospin texture is presented in Figs.~\ref{fig: extra}(e,f). The arrows represent in-plane components  $(S_x,S_y)$, while the colors represent the out-of-plane component $S_z$. In the $\Delta=0$ case, the pseudospin textures of the upper and lower eigenstates have opposite in-plane components $(S_x,S_y)$ while their out-of-plane components $S_z$ are aligned.  The latter highlights the non-orthogonality of the eigenstates\cite{su2021} in this non-Hermitian system, which is maximized at the exceptional points.}

In the following, we use the two methods of calculating the QGT: (1) directly from the eigenstates as prescribed theory Eqs.~(\ref{eq: RR QGT}, \ref{eq: LR QGT}), and (2) from the pseudospin texture to emulate the proposed experimental method.

\subsection{Right-Right Quantum Geometric Tensor}

The components of the RR QGT, derived from the eigenstates of the exciton-polariton Hamiltonian $H_{pol}$ using Eq.~(\ref{eq: RR QGT}), take the forms:
\red{
\begin{equation}\label{eq: QRR polariton}
    \begin{split}
        g_{\pm,xx}^{RR}=&\frac{4|\tilde\beta|^2\left(k_y^2|h_+|^2|\tilde\alpha-\tilde\beta k^2|^2-2k_y\operatorname{Im}[G_{\pm,x}]+\Delta^2 k^2|\Delta\pm\lambda|^2\right)}{|\lambda|^2\left(|h_+|^2+|\Delta\pm \lambda|^2\right)^2}\\
        g_{\pm,xy}^{RR}=&\frac{4|\tilde\beta|^2\left(k_xk_y|h_+|^2\left(|\tilde\alpha|^2-|\tilde\beta|^2k^4\right)+\operatorname{Re}\left[k_y G_{\pm,x}^*+i k_x G_{\pm,y}\right]\right)}{|\lambda|^2\left(|h_+|^2+|\Delta\pm \lambda|^2\right)^2}\\
        g_{\pm,yy}^{RR}=&\frac{4|\tilde\beta|^2\left(k_x^2|h_+|^2|\tilde\alpha+\tilde\beta k^2|^2-2k_x\operatorname{Im}[G_{\pm,y}]+\Delta^2 k^2|\Delta\pm\lambda|^2\right)}{|\lambda|^2\left(|h_+|^2+|\Delta\pm \lambda|^2\right)^2}\\
        \Omega_{\pm}^{z,RR}=&\frac{8|\tilde\beta|^2\left(2k_xk_yk^2|h_+|^2(b\alpha-a\beta)-\operatorname{Im}\left[k_y G_{\pm,x}^*+i k_x G_{\pm,y}\right]-\Delta^2k^2|\Delta\pm\lambda|^2\right)}{|\lambda|^2\left(|h_+|^2+|\Delta\pm \lambda|^2\right)^2}
    \end{split}
\end{equation}
where we denote $G_{\pm,x}=h_+(\tilde\alpha-\tilde\beta k^2)(k_x-ik_y)\Delta(\Delta\pm\lambda^*)$, $G_{\pm,y}=h_+(\tilde\alpha+\tilde\beta k^2)(k_x-ik_y)\Delta(\Delta\pm\lambda^*)$.}
Interestingly, the $|\lambda|^2$ in the denominator ensures that all components of the RR QGT in exciton polaritons diverge at the spectral degeneracies, where $\lambda = 0$, unlike in previous work on non-Hermitian SSH models~\cite{ye2023}. At $\Delta=0$, the divergence is $\sim k^{-1}$, in agreement with previous results following a similar approach~\cite{liao2021}. 

\begin{figure}[h]
    \centering
    \includegraphics[width=0.8\textwidth]{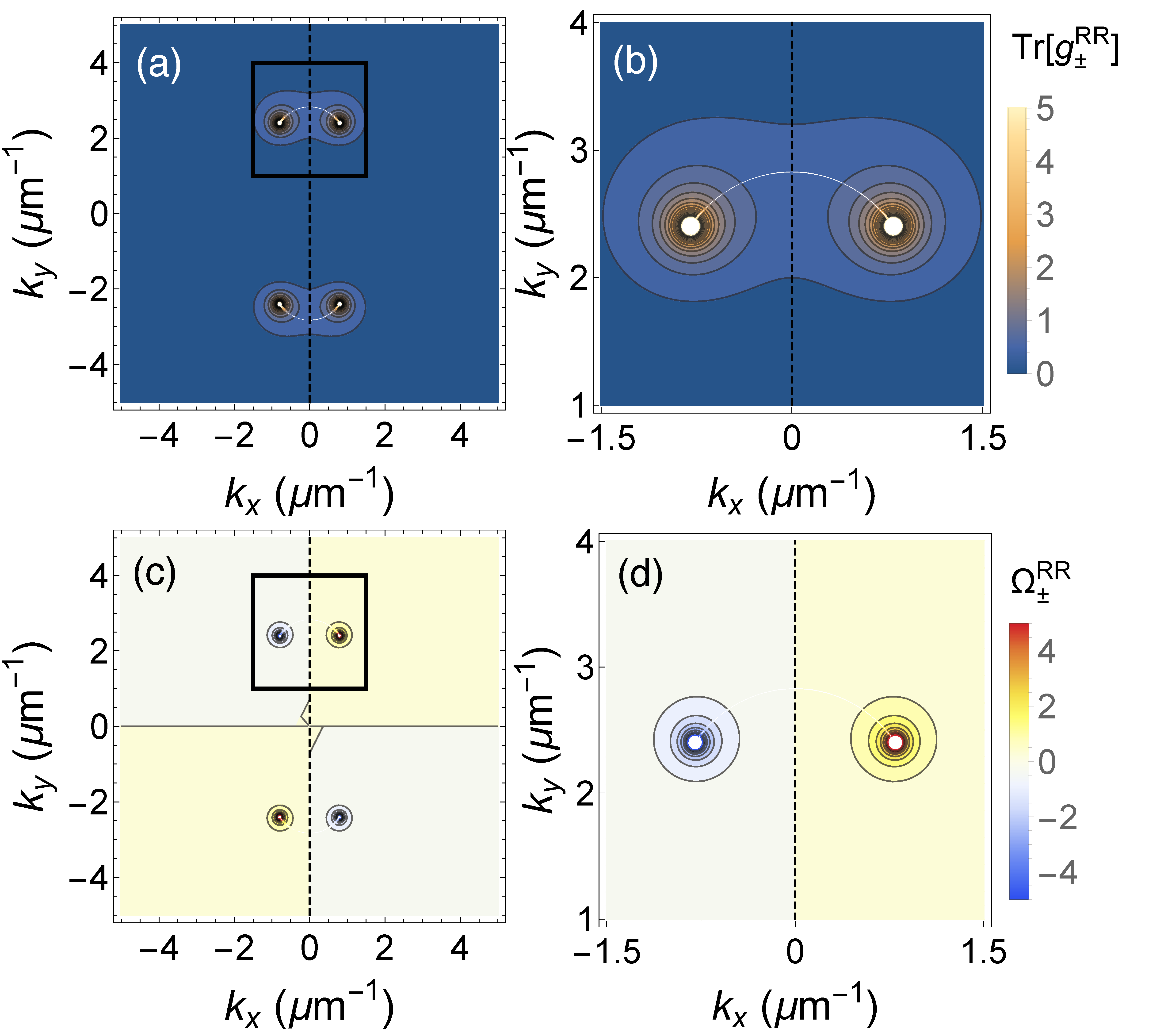}
    \caption{The trace of (a) the RR QMT and (c) the RR Berry curvature calculated from the eigenstates (left-half panel) and from the pseudospin components (right-half panel) at $\Delta=0$ eV. Also showing the zoom-in near the exceptional points of $\operatorname{Tr}[g_\pm^{RR}]$ and $\Omega_\pm^{z,RR}$ in (b) and (d), respectively.}
    \label{fig: 1}
\end{figure}

\begin{figure}[h]
    \centering
    \includegraphics[width=0.95\textwidth]{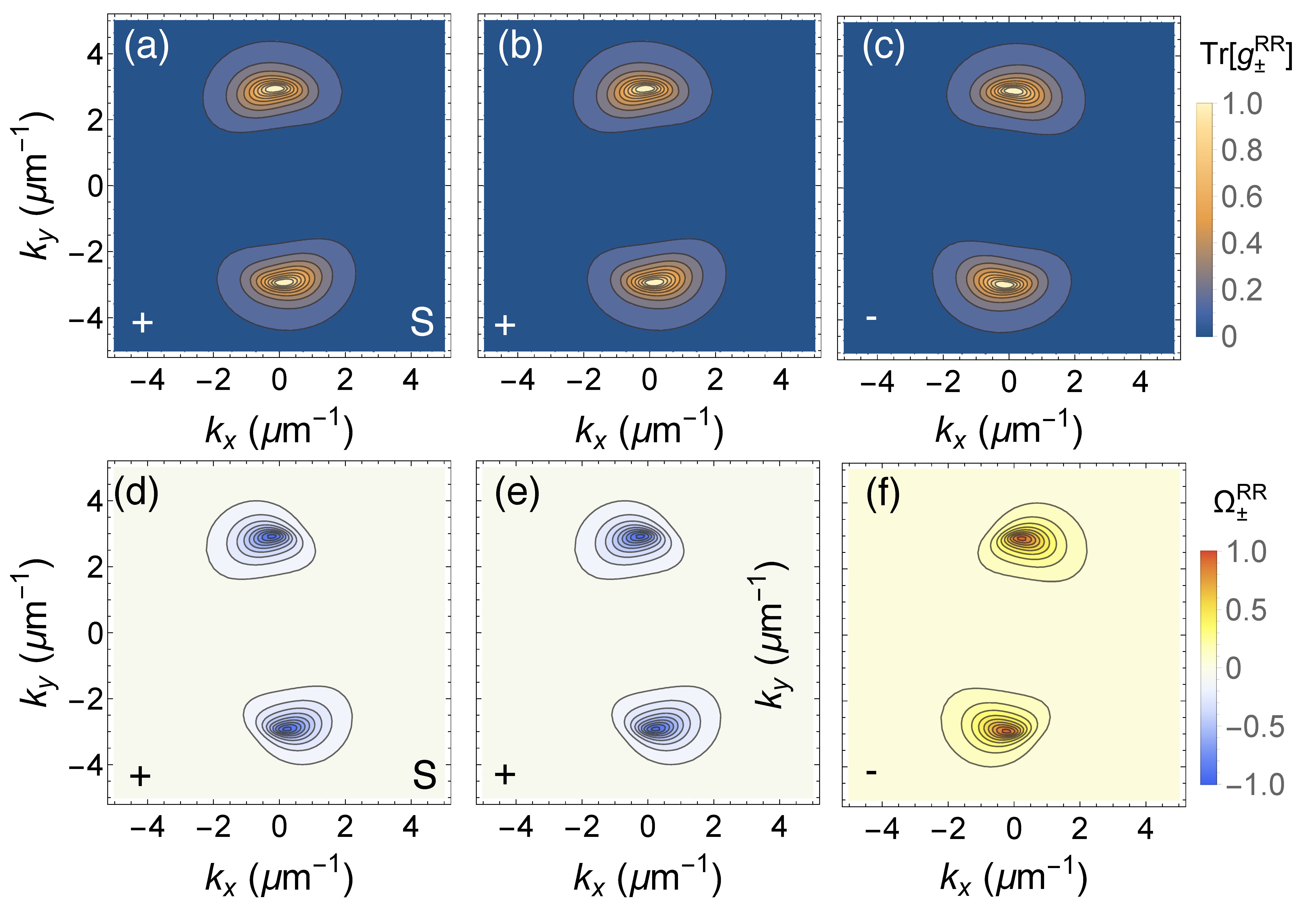}
    \caption{The traces of the right-right quantum geometric tensors of (b) the upper and (c) the lower bands. Also showing right-right Berry curvatures of (e) the upper and (f) the lower bands in a gapped phase with $\Delta=0.0075$ eV, slightly larger than $|\Delta_c|=0.00599$ eV (see Supplemental Document). The results calculated from the eigenstates (b,e) show excellent agreement with the ones calculated from the pseudospins (a,d).}
    \label{fig: 2}
\end{figure}

Figure~\ref{fig: 1} shows the trace of the RR QMT and the RR Berry curvature calculated using Eqs.~(\ref{eq: QRR polariton}), which clearly shows the divergence of the components at the exceptional points. Note that the divergence of the Berry curvature is opposite for exceptional points in the same pair, which is reminiscent of their opposing  spectral winding~\cite{su2021}. Note also that, unlike the Hermitian case~\cite{bleu2018}, the divergence of the RR QMT components will persist for small values of $\Delta$ as long as the exceptional points persist.

\red{For larger $\Delta$, such that the exceptional points are annihilated and the gap is opened, the components of the RR QMT take on finite values as shown in Fig.~\ref{fig: 2}. There are two striking features compared to the $\Delta=0$ case. Firstly, the components are asymmetrically distributed in momentum space. Secondly, the upper and lower bands are not exactly equal nor opposite to each other, violating the identities for the Hermitian limit [see Eqs.~(\ref{eq: two level}) ].  These features arise from the $\Delta\pm\lambda^{(*)}$ factors in Eqs.~(\ref{eq: QRR polariton}), which originated from the} non-orthogonality of the right-eigenstates \cite{sternheim1972,zhang2019}.

\red{The $\Delta=0$ case features a symmetrical distribution [see Fig.~\ref{fig: 1}] with components from the opposite bands aligned with each other. However, it is important to highlight that the Berry curvatures of the two bands are the same, i.e., $\Omega_+^{RR}=\Omega_-^{RR}$, instead of having the opposite signs [see Eqs.~(\ref{eq: two level})]. This is reminiscent of the $S_z$ components of the right eigenstates, which are pointing in the same direction due to the non-orthogonality of the eigenstates~\cite{su2021} [see Fig.~\ref{fig: extra}(e,f)].}

To emulate the proposed experiment, we also calculate the components of the RR QMT from the polariton pseudospins using Eqs.~(\ref{eq: QRR S}). The resulting distributions are plotted in Fig.~\ref{fig: 1}, together with those derived directly from the eigenstates. We can see that the quantities calculated using these two different methods agree very well, which confirms the applicability of the formalism presented in Section~\ref{sec: qgt s} for non-Hermitian systems.

\subsection{Left-Right Quantum Geometric Tensor}

\begin{figure}[h]
    \centering
    \includegraphics[width=0.8\textwidth]{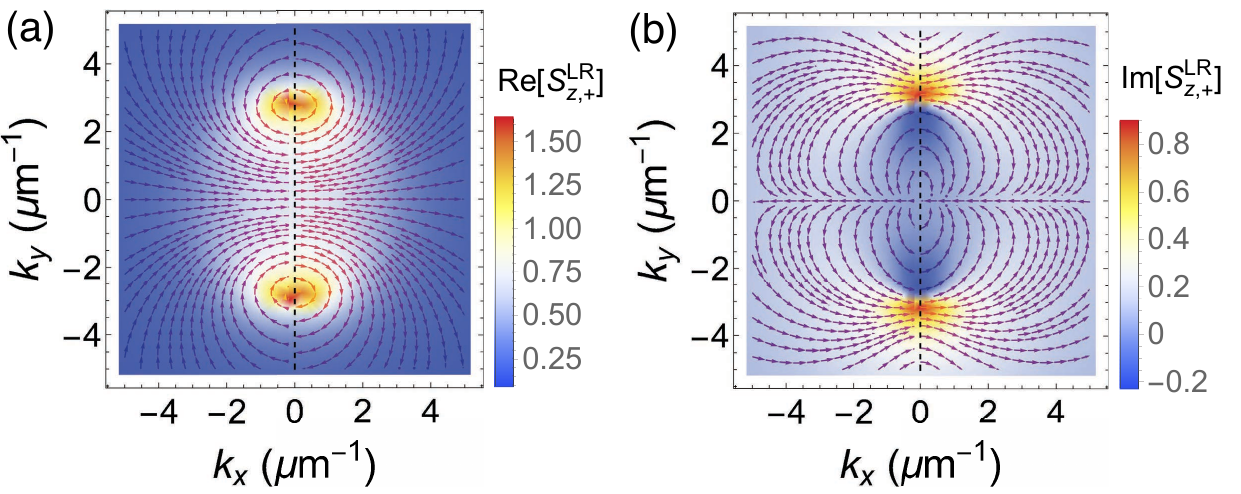}
    \caption{(a) Real part and (d) imaginary parts of $\mathbf{S}^{LR}_+$ calculated from the eigenstates (left-half panel) and the pseudospins $\mathbf{S}^R_\pm$ (right-half panel), respectively. The color denotes the out-of-plane components $S^{LR}_{z,+}$ while the arrows denote the in-plane components $(S^{LR}_{x,+},S^{LR}_{y,+})$. To open the a gap, $\Delta=0.0075$ eV was chosen.}
    \label{fig: extra2}
\end{figure}

As noted in Section~\ref{sec: qgt s}, the components of $\mathbf{S}^{LR}_\pm$ can be constructed from the pseudospins of the right eigenstates, which can be experimentally measured. \red{The $\mathbf{S}^{LR}_\pm$ calculated from both the eigenstates and from the experimentally measurable pseudospin $\mathbf{S}^{R}_\pm$ is shown in Fig.~\ref{fig: extra2}, showing excellent agreement.} The latter method provides a way to experimentally measure the components of the LR QGT, $Q^{LR}_{\pm,\mu\nu}$, in an exciton-polariton system.

\begin{figure}[h!]
    \centering
    \includegraphics[width=0.8\textwidth]{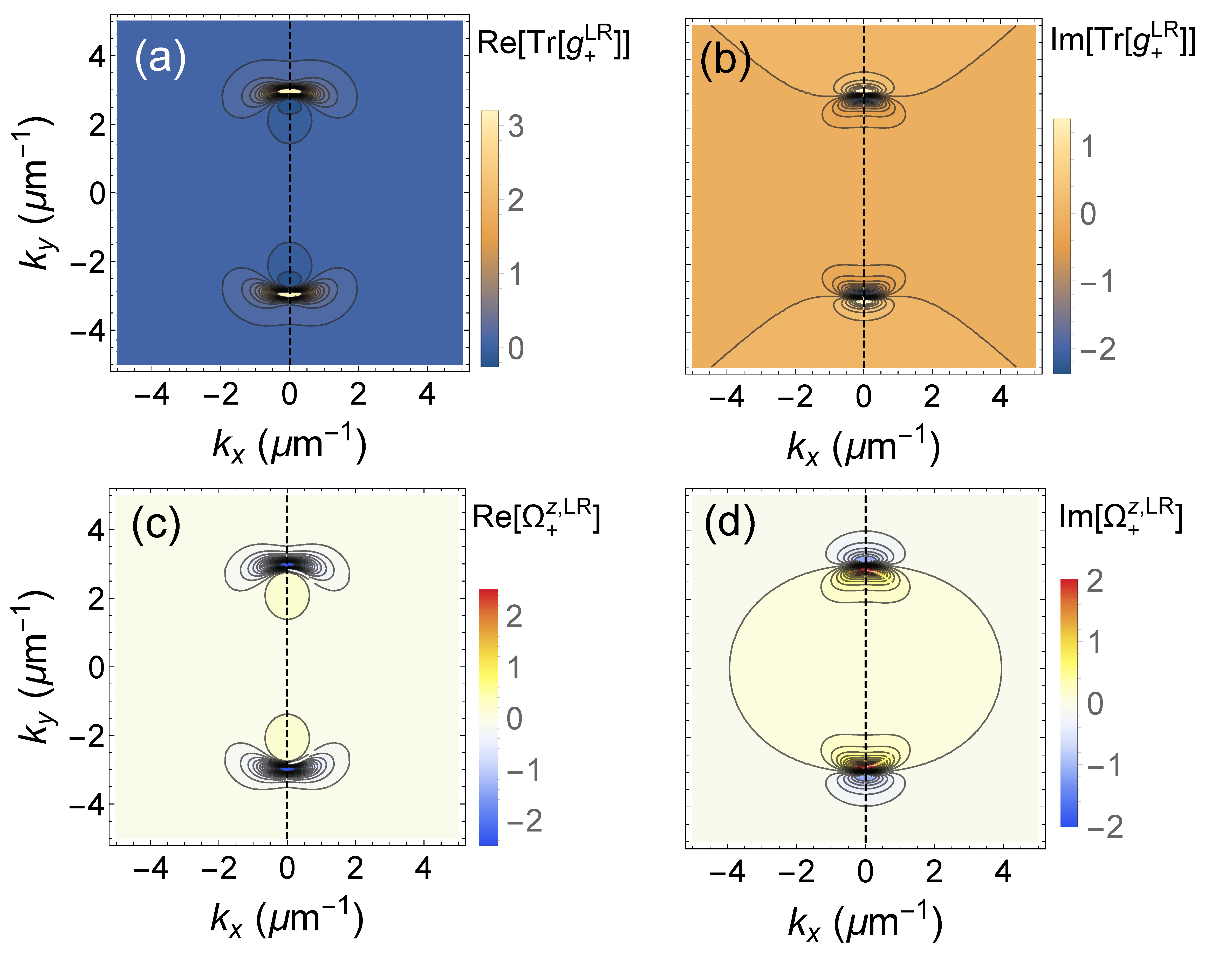}
    \caption{The real and imaginary parts of the trace of the (a), (b) LR QMT and the (c), (d) LR Berry curvature of the upper band, respectively. The left-half of each panel is calculated from the eigenstates and the right half is calculated from the pseudospins. The Zeeman splitting is large enough ($\Delta=0.0075$ eV) such that the exciton-polariton bands are the gapped.}
    \label{fig: 3}
\end{figure}

The components of the LR QGT in the exciton-polariton system with the Hamiltonian Eq.~(\ref{eq: polariton H}) take the following form: 
\begin{equation}\label{eq: QNH}
    \begin{split}
        g^{LR}_{\pm,xx}=&\frac{\tilde\beta^2\left[(\tilde\alpha-\tilde\beta k^2)^2k_y^2+\Delta^2k^2\right]}{\lambda^4} \\
         g^{LR}_{\pm,xy}=&\frac{\tilde\beta^2k_x k_y\left(\tilde\alpha^2-\tilde\beta^2k^4\right)}{\lambda^4}\\
         g^{LR}_{\pm,yy}=&\frac{\tilde\beta^2\left[(\tilde\alpha+\tilde\beta k^2)^2k_x^2+\Delta^2k^2\right]}{\lambda^4}\\
        \Omega^{z,LR}_\pm=&\mp\frac{2k^2\tilde\beta^2\Delta}{\lambda^3}.
    \end{split}
\end{equation}
\red{To visualize these in momentum space, we plot them in Fig.~\ref{fig: 3} and compare with those derived from the} pseudospins using Eq.~(\ref{eq: nh qgt}). The plots clearly show that the formalism presented in Section~\ref{sec: qgt s} works well for the LR QGT.

Unlike in the RR QGT, the identities $g_{+,\mu\nu}=g_{-,\mu\nu}$, $\Omega_{+,\mu\nu}=-\Omega_{-,\mu\nu}$ were recovered for the LR QGT, since the components of the LR QGT can be expressed as
\begin{equation}
    \begin{split}
        g^{LR}_{n,\mu\nu}&=\frac{1}{2}\sum_{m\neq n}\left[\frac{\langle\psi_m^L|\partial_\mu H|\psi_n^R\rangle\langle\psi_n^L|\partial_\nu H|\psi_m^R\rangle}{(E_m-E_n)^2}+\frac{\langle\psi_m^L|\partial_\nu H|\psi_n^R\rangle\langle\psi_n^L|\partial_\mu H|\psi_m^R\rangle}{(E_m-E_n)^2}\right]\\
        \Omega^{LR}_{n,\mu\nu}&=i\sum_{m\neq n}\left[\frac{\langle\psi_m^L|\partial_\mu H|\psi_n^R\rangle\langle\psi_n^L|\partial_\nu H|\psi_m^R\rangle}{(E_m-E_n)^2}-\frac{\langle\psi_m^L|\partial_\nu H|\psi_n^R\rangle\langle\psi_n^L|\partial_\mu H|\psi_m^R\rangle}{(E_m-E_n)^2}\right]
    \end{split}
\end{equation}
similarly to the Hermitian case in Eq.~(\ref{eq: qgt pert}) \cite{sternheim1972,zhang2019}. Additionally, the symmetry of the distributions in momentum space is restored (cf. Fig.~\ref{fig: 2} ).

Compared to the components of $Q^{RR}_{\pm,\mu\nu}$, the components of $Q^{LR}_{\pm,\mu\nu}$ take very similar forms to the ones in the Hermitian limit \cite{bleu2018} but with $(\alpha,\beta)\rightarrow(\tilde\alpha,\tilde\beta)$. Similarly, one can also recover the RL $g^{RL}_{\pm,\mu\nu}$ and $\Omega^{z,RL}_\pm$ by taking the Hermitian results and replacing $(\alpha,\beta)\rightarrow(\tilde\alpha^*,\tilde\beta^*)$.

We also note that, unlike the RR QGT, $Q^{RR}_{\pm,\mu\nu}$ at $\Delta=0$, the quantum metric $g^{LR}_{\pm,\mu\nu}$ diverges as $\sim k^{-2}$ at each exceptional point while the Berry curvature $\Omega^{z,LR}_{\pm}$ diverges as $\sim k^{-3/2}$, in agreement with previous work~\cite{brody2013}. Furthermore, although both $\Omega^{z,LR}_\pm$ and $\Omega^{z,RR}_\pm$ diverge at the exceptional points, the LR Berry curvature $\Omega^{z,LR}=0$ elsewhere but the RR Berry curvature $\Omega^{z,RR}_\pm$ is finite. The delta-function behavior of the former is reminiscent of the Berry curvature near Dirac points in the Hermitian limit~\cite{bleu2018}. If the finite value of the RR Berry curvature is physically relevant, we expect it to influence the wave packet dynamics via the anomalous Hall drift, which we explore in the following section.

\section{Anomalous Hall Drift and the Role of non-Hermitian Berry Connection}\label{sec: WP}
\red{In this Section, we analyze the anomalous Hall drift arising from the Berry curvature by studying the wave-packet dynamics in the exciton-polariton model Eq.~(\ref{eq: polariton H}). This will guide experiments that could allow to directly measure the QGT components and to determine their effect on another observable, e.g., the trajectory of wave-packet center of mass (COM)~\cite{gianfrate2020}. In Hermitian systems, applying an external field perpendicular to the Berry curvature results in an anomalous Hall drift. In non-Hermitian systems, a more complex anomalous drift occurs but other effects also come into play such as inter-band transitions and self-acceleration~\cite{silberstein2020,longhi2022,solnyshkov2021,hu2022,yuce2017}, even in the absence of the external field. Here, we attempt to untangle these effects to highlight the direct effect of the Berry curvature.}

\red{First, we consider the case of an extremely narrow wave packet in momentum space, i.e., $\langle\psi|\psi\rangle\approx\delta(\mathbf{k}-\mathbf{k}_c)$}. Previous works \cite{xu2017,silberstein2020,wang2022} have shown that the dynamics of the wave-packet COM in a non-Hermitian system can be approximated by a pair of semi-classical equations of motion as follows:
\begin{equation}\label{eq: nh ahe}
    \begin{split}
        \hbar\Dot{\mathbf{k}_c}&=\mathbf{F}\\
        \hbar\Dot{\mathbf{r}_c}&=\nabla_\mathbf{k}\operatorname{Re}[E_\pm+\mathbf{F}\cdot(\mathbf{A}_\pm^{RR}-\mathbf{A}_\pm^{LR})]-\mathbf{F}\times\Omega_\pm^{z,RR}.
    \end{split}
\end{equation}
where $\mathbf{k}_c$, $\mathbf{r}_c$ are the wave-packet COM momentum and position, respectively, $\mathbf{F}$ represents an external force arising from a real-space potential $\mathbf{F}=-\nabla_{\mathbf{r}}U(\mathbf{r})$. The last term is the anomalous Hall drift due to the RR Berry curvature $\Omega_\pm^{z,RR}$ while the first term contains the group velocity and the non-Hermitian field-induced correction~\cite{silberstein2020}. Here, $\mathbf{A}_\pm^{RR}$ and $\mathbf{A}_\pm^{LR}$ are the RR and LR Berry connections \cite{silberstein2020}, respectively defined as:
\begin{equation}
\begin{split}
    \mathbf{A}_\pm^{RR}&=\frac{\langle\psi_\pm^R|i\nabla_\mathbf{k}\psi_\pm^R\rangle}{\langle\psi_\pm^R|\psi_\pm^R\rangle}\\
    \mathbf{A}_\pm^{LR}&=\langle\psi_\pm^L|i\nabla_\mathbf{k}\psi_\pm^R\rangle.
\end{split}
\end{equation}
with normalization $\langle\psi_n^L|\psi_n^R\rangle=1$, which are necessary to derive Eq.~(\ref{eq: nh ahe}).  The $(\mathbf{A}_\pm^{RR}-\mathbf{A}_\pm^{LR})$ term, which we call the non-Hermitian Berry connection, can be measured experimentally, similarly to the components of the non-Hermitian QGT (see Supplemental Document).

\red{The second line in Eq.~\ref{eq: nh ahe} clearly shows that, in addition to the anomalous drift due to the Berry curvature, the Berry connection now plays a direct role resulting in an additional anomalous drift. Note that $(\mathbf{A}_\pm^{RR}-\mathbf{A}_\pm^{LR})=0$ in the Hermitian limit. Hence, in addition to the Berry curvature, the non-Hermitian Berry connection (which can also be measured) should be considered in analyzing anomalous drifts.
}

\begin{figure}[h]
    \centering
    \includegraphics[width=0.9\textwidth]{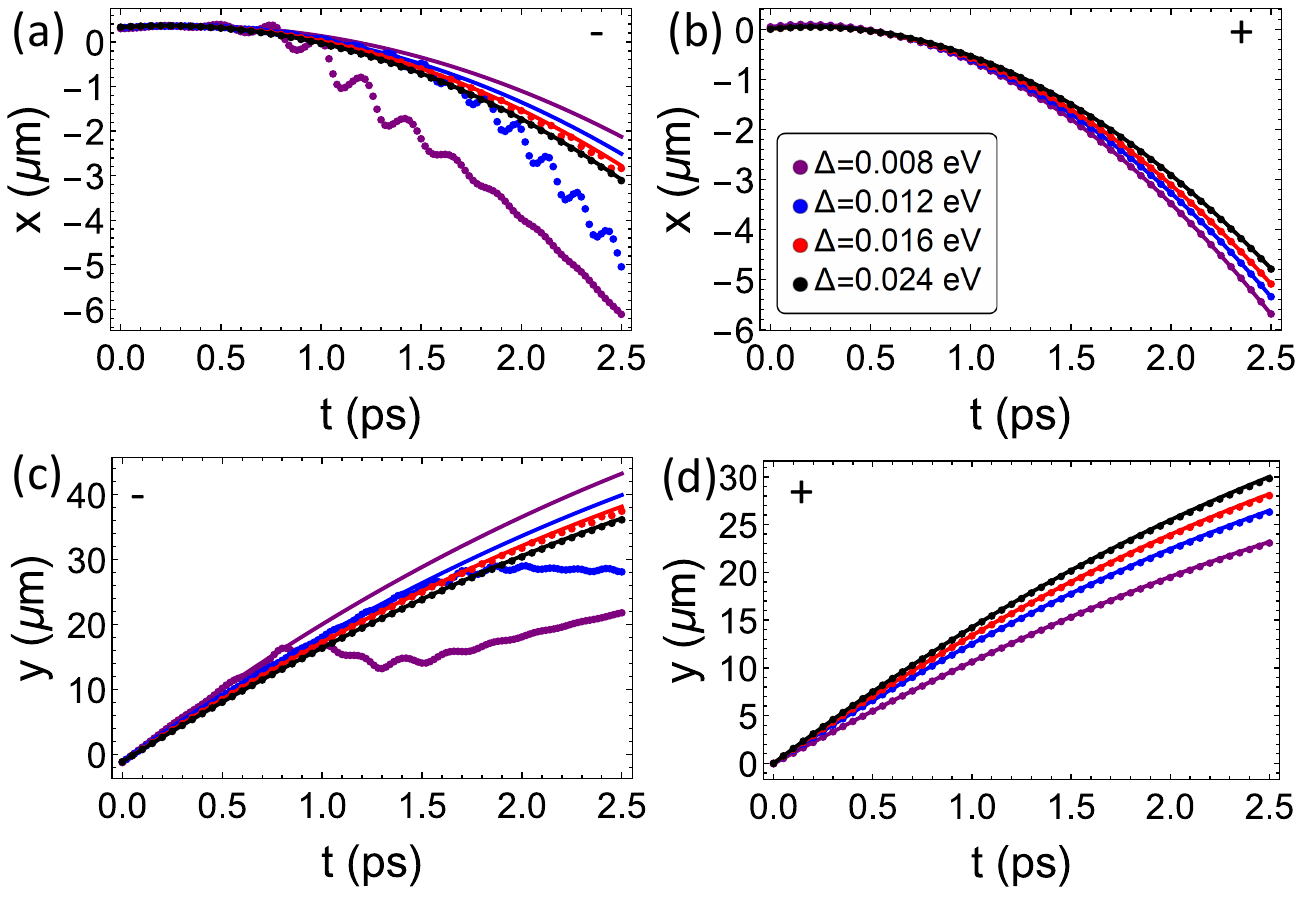}
    \caption{The trajectories of wave packets in the lower (a and c) and upper (b and d) eigenstates with different values of $\Delta$, which are chosen to be larger than $|\Delta_c|=0.0059$~eV. The dots represent numerical results while the lines represent analytic results derived using Eq.~(\ref{eq: nh ahe}).}
    \label{fig: wp0}
\end{figure}

We tested the semi-classical equation of motion Eq.~(\ref{eq: nh ahe}) by simulating a narrow wave packet initially prepared in the upper eigenstate and with a width of $w=0.01\mu$m$^{-1}$. \red{Specifically, we construct the time-evolution operator from the Schrodinger equation to propagate the initial wave packet forward in time~\cite{hu2022} , as follows
\begin{equation}\label{eq: gaussian}
    |\psi(\mathbf{k},t)\rangle=e^{-i(\mathbf{H(\mathbf{k})}-\mathbf{F}\cdot\mathbf{r})t/\hbar}e^{-\frac{(\mathbf{k}-\mathbf{k}_0)^2}{4w^2}}e^{i\mathbf{k}\cdot\mathbf{r}_0}|\psi^R_\pm(\mathbf{k})\rangle.
\end{equation}
}The wave packet has initial COM momentum of  $\mathbf{k}_c=(0.05,2.5)\mu$m$^{-1}$, where the two bands are close to each other, and initial COM position of $\mathbf{r}_c=(0,0)\mu$m with an external force $\mathbf{F}=(-1.5\times10^{-4},-3\times10^{-4})$~eV/$\mu$m. We obtain the numerical results of wave-packet trajectories by using split-step Fourier methods \cite{taha2005,weideman1986}. 

Our numerical results show great agreement with analytic results for \blue{the wave packets initially in the upper eigenstates for four} different values of $\Delta$, as shown in \blue{Figs.~\ref{fig: wp0}(b,d)}. However, for the wave packets initially prepared in the lower eigenstate, the analytic equations only fit well with the numerical simulations for larger $\Delta=0.016$eV [see Figs.~\ref{fig: wp0}(a,c)]. The oscillation we observe arises from the Zitterbewengung effect due to the inter-band mixing \cite{leblanc2021,sedov2018}. On the path of the wave packet in momentum space, the upper eigenstate has the lower decay rate or larger imaginary part. Therefore, the wave packet initially prepared in the lower eigenstate evolves towards the upper eigenstate due to the inter-band transition, which is neglected in the semi-classical approximation~\cite{hu2022,silberstein2020,pan2020}.

\red{Therefore, to minimize effects due to inter-band transitions, it is important to work in the regime where the two bands are well separated and more importantly, to initially prepare the wave packet in the band with the smallest decay rate.}

\begin{figure}[h]
    \centering
    \includegraphics[width=0.9\textwidth]{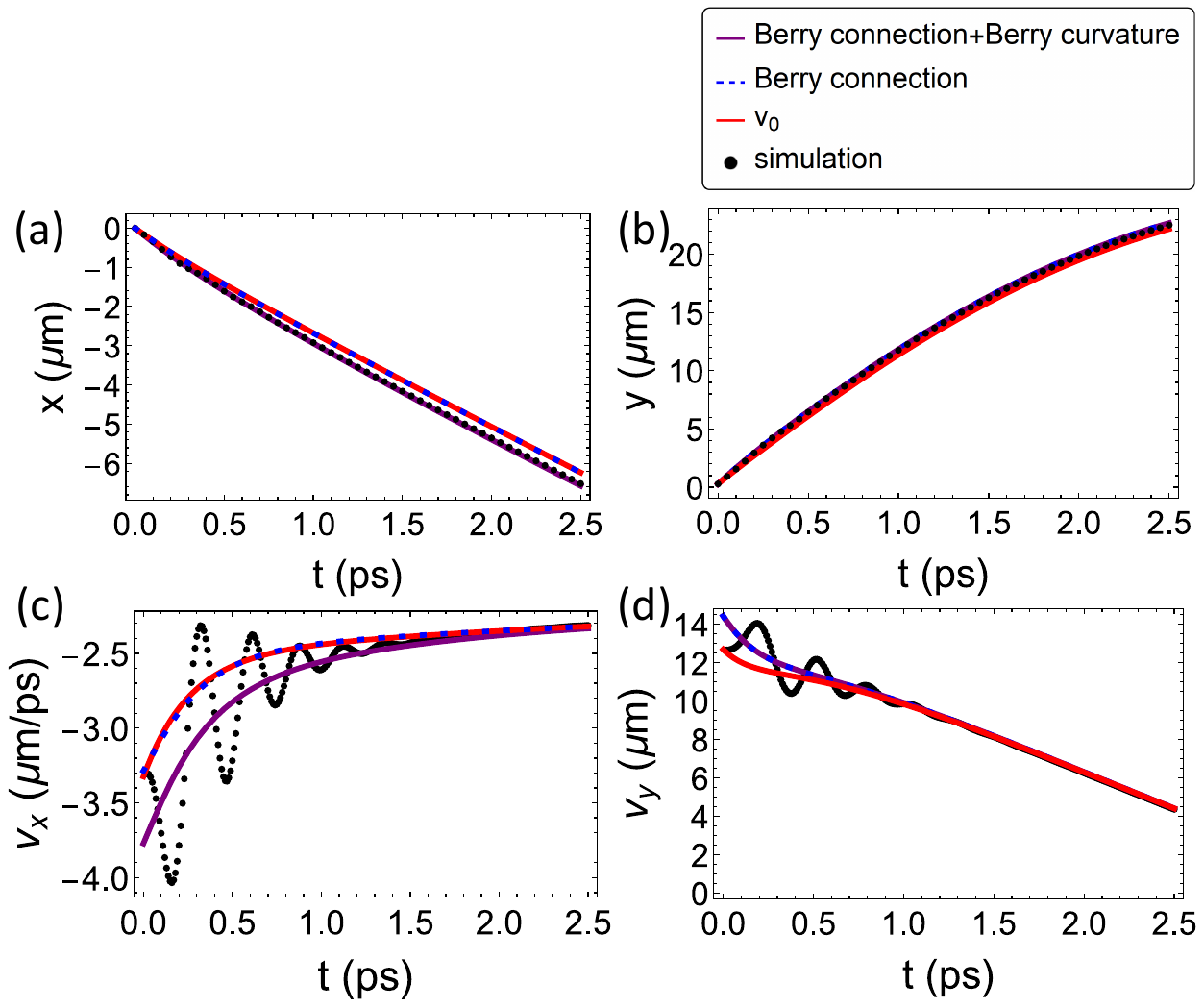}
    \caption{(a,b) Position and (c,d) group velocity of a simulated wave packet in a gapped polariton system with $\Delta=0.008$eV, initial COM momentum $\mathbf{k}_c=(-0.25,2.8)\mu$m$^{-1}$, where the magnitude of the RR Berry curvature is large, and under a constant force $\mathbf{F}=(0,-5\times10^{-4})$eV/$\mu$m. The red, blue and purple lines represent the contribution from $\mathbf{v}_0=\nabla_\mathbf{k}\operatorname{Re}[E_\pm]$ only, the contribution from both $\mathbf{v}_0$ and $\operatorname{Re}[\mathbf{A}^{RR}_\pm-\mathbf{A}^{LR}_\pm]$, and the contribution from all of $\mathbf{v}_0$, $\operatorname{Re}[\mathbf{A}^{RR}_\pm-\mathbf{A}^{LR}_\pm]$ and $\operatorname{Re}\Omega_\pm^{z,RR}$, respectively. Numerical data are plotted with black dots}
    \label{fig: wpb}
\end{figure}

To directly observe the anomalous Hall drift due to the RR Berry curvature, we simulated the dynamics of a wave packet with initial width of $w=0.01\mu$m$^{-1}$, and initial COM momentum of $\mathbf{k}_c=(-0.25,2.8)\mu$m$^{-1}$, close to one of the local maxima of $|\Omega_+^{RR}|$, and initial COM position of $\mathbf{r}_c=(0,0)\mu$m. We chose a Zeeman splitting of $\Delta=0.008$ eV to open a gap, and apply a constant force $\mathbf{F}=(0,-5\times10^{-4})$~eV/$\mu$m along the $y$-direction in order to observe the anomalous Hall drift in the $x$-direction [see Fig.~\ref{fig: wpb}]. Here, we also observe oscillation in the group velocities in Figs.~\ref{fig: wpb}(c,d), which arises from the Zitterbewegung effect \cite{sedov2018,leblanc2021} since the wave packet was initially prepared at where the two bands are close. However, since the wave packet was initially prepared in the upper eigenstate, which has the larger $\operatorname{Im}E$, the oscillation in the COM position is much smaller than in Figs.~\ref{fig: wp0}(a,c).

\red{At first look, the RR Berry curvature seems to be the main relevant component of the QGT at play here. However, we can rewrite Eq.~(\ref{eq: nh ahe}) as:
\begin{equation}\label{eq: nh ahe2}
\hbar\Dot{\mathbf{r}_c}=\nabla_\mathbf{k}\operatorname{Re}[E_\pm]+(\mathbf{F}\cdot\nabla_\mathbf{k})\operatorname{Re}[\mathbf{A}_\pm^{RR}-\mathbf{A}_\pm^{LR}]-\mathbf{F}\times\operatorname{Re}[\Omega_\pm^{z,LR}]
\end{equation}
where it now looks like the LR Berry curvature is the main component but with a modified correction due to the non-Hermitian Berry connection. Hence, the choice of representation is arbitrary and both Berry curvature and Berry connections should be considered in analyzing field-induced anomalous drifts.

This indicates that, in non-Hermitian systems, the left and right eigenstates are needed when deriving relevant geometric quantities like the QGT components and the Berry connection. This is further supported by the fact that the numerical simulations and the derivation of Eqs.~(\ref{eq: nh ahe}) and~(\ref{eq: nh ahe2}) did not invoke any biorthogonal quantum mechanics~\cite{brody2014}, yet the LR Berry curvature and LR Berry connection naturally appear.}

\red{Another interesting implication of Eq.~(\ref{eq: nh ahe}) and~(\ref{eq: nh ahe2}) is the possibility of observing an anomalous Hall drift without Zeeman splitting. In the Hermitian limit, the Berry curvature is zero when $\Delta=0$~\cite{bleu2018}, but for the system considered here, the RR Berry curvature is finite around the exceptional points [see Eq.~(\ref{eq: RR QGT}) and Fig.~\ref{fig: 1}(c,d)], and hence will induce an anomalous drift. Alternatively, the LR Berry curvature is zero outside the exceptional point so the non-Hermitian Berry connection plays a dominant role in the anomalous drift. However, as discussed above, Eq.~(\ref{eq: nh ahe}) will fail near the exceptional points since the bands are energetically close to each other leading to inter-band transitions and it will be difficult to isolate effects due to the Berry curvature. Analyzing wave-packet dynamics including inter-band transitions will be the subject of future work.
}

\subsection{Finite Size Correction and Self-Acceleration}
\red{Another effect in non-Hermitian systems is self-acceleration due to the finite width of the wave packets~\cite{silberstein2020,longhi2022,solnyshkov2021,hu2022,yuce2017} even in the absence of an external force.}
Since experimental wave packets will always have a non-zero width in momentum space, it is important to include finite-size effects in analyzing wave-packet dynamics to uncover any anomalous Hall drift.

A semi-classical equation of motion including finite-size effects in the absence of an external force was previously derived in Ref. \cite{silberstein2020}.  Here, we extend the formalism to include both the external force and self-acceleration while assuming that the single-band approximation holds \cite{hu2022,longhi2022,solnyshkov2021,yuce2017}. The resulting equations, up to the second order in the wave-packet widths, along the $i$-th direction are
\begin{equation}\label{eq: finite sigma}
    \begin{split}
        \hbar(\Dot{\mathbf{k}}_c)_i=&F_i + 2w_i^2\partial_{k_i}\operatorname{Im}[E_\pm+\mathbf{F}\cdot(\mathbf{A}_\pm^{RR}-\mathbf{A}_\pm^{LR})]\\
        \hbar(\Dot{\mathbf{r}}_c)_i=&\partial_{k_i}\operatorname{Re}[E_\pm+\mathbf{F}\cdot(\mathbf{A}_\pm^{RR}-\mathbf{A}_\pm^{LR})]-(\mathbf{F}\times \Omega_{\pm}^{RR})_i\\
        &+\sum_{j}\Bigg( \frac{w_j^2}{2}\partial_{k_j}^2\left(\partial_{k_i}\operatorname{Re}\left[E_\pm+\mathbf{F}\cdot(\mathbf{A}_\pm^{RR}-\mathbf{A}_\pm^{LR})\right]-(\mathbf{F}\times \Omega_{\pm}^{RR})_i\right)\\
        &+2w_j^2\left[\partial_{k_j}\operatorname{Im}\left[E_\pm+\mathbf{F}\cdot(\mathbf{A}_\pm^{RR}-\mathbf{A}_\pm^{LR})\right]\partial_{k_j}\left(\operatorname{Re}\left[(\mathbf{A}_\pm^{RR})_i\right]-\partial_{k_i}\varphi\right)\right]\Bigg)
    \end{split}
\end{equation}
where $\varphi$ is the complex phase of the wave packet and $w_j$ is the wave-packet width along the $j$-th direction in momentum space. The latter parameter can be computed as $$ w_j^2=\frac{\int\left(k_j-(\mathbf{k}_c)_j\right)^2\langle\psi(\mathbf{k},t)|\psi(\mathbf{k},t)\rangle d^2\mathbf{k}}{\int\langle\psi(\mathbf{k},t)|\psi(\mathbf{k},t)\rangle d^2\mathbf{k}}. $$

\begin{figure}[h]
    \centering
    \includegraphics[width=0.9\textwidth]{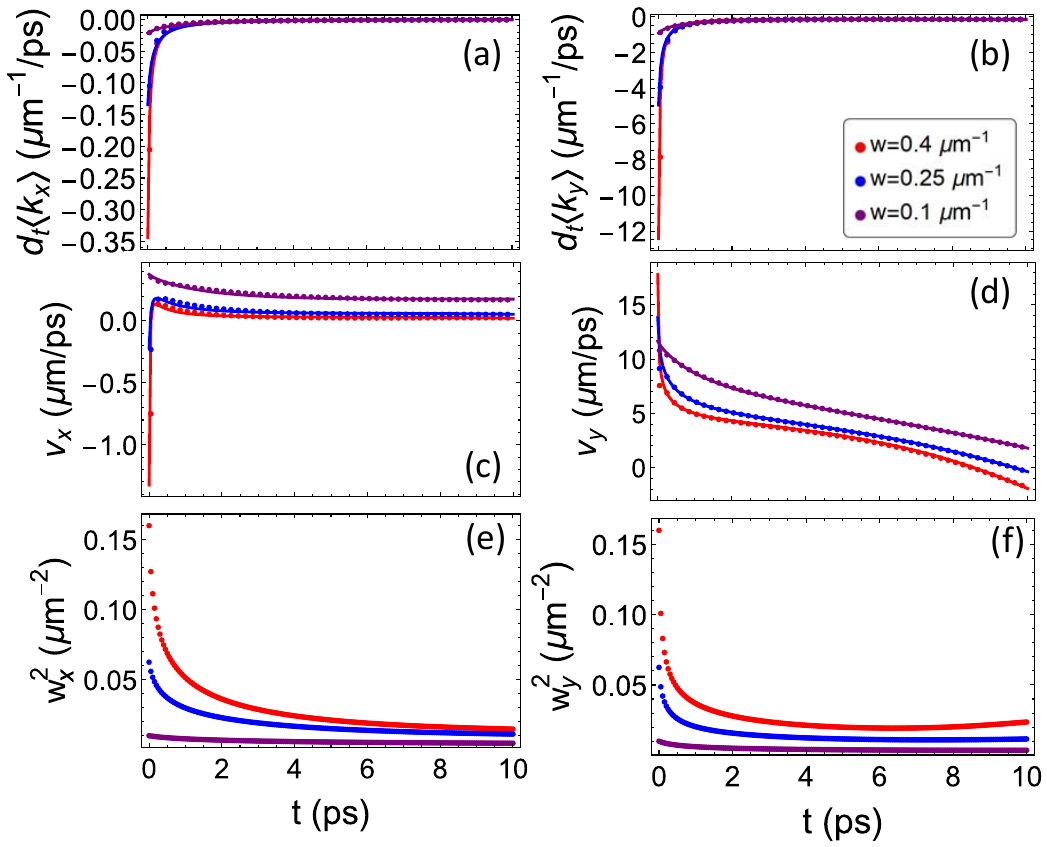}
    \caption{(a,b) Time derivatives of the COM momenta and (c,d) the group velocities computed numerically (dots) and semi-analytically using Eq.~(\ref{eq: finite sigma}) (lines). (e,f) The numerically extracted wave-packet widths. The values of $w$ in the legend correspond to the initial wave-packet widths.}
    \label{fig: wp sigma}
\end{figure}

To test the approximation in Eq.~(\ref{eq: finite sigma}), we simulate the dynamics of Gaussian wave packets initially prepared in the upper eigenstates with different initial widths $w$, and initial COM at $\mathbf{k}_c=(0.05,2.5)\mu$m$^{-1}$, $\mathbf{r}_c=(0,0)\mu$m and an external force  $\mathbf{F}=(0,-10^{-4})$~eV/$\mu$m. The complex phase $\varphi$ from Eq.~(\ref{eq: finite sigma}) is difficult to compute but we assume that the terms proportional to the external force in Eq.~(\ref{eq: gaussian}) are small and approximate the terms as $\partial_{k_i}\varphi\approx-\partial_{k_i}\operatorname{Re}E_+t/\hbar+(\mathbf{r}_0)_i$ where $\mathbf{r}_0$ disappear in the derivatives. Note that in contrast with Hermitian systems, the wave-packet size (in momentum space) changes as it evolves, which makes it difficult to compute Eq.~(\ref{eq: finite sigma}) completely analytically. We therefore adopt a semi-analytic approach and numerically extract $w_j^2$ [see Figs.~\ref{fig: wp sigma}(e,f)], then use it to compute the time derivatives of the COM momenta [see Figs.~\ref{fig: wp sigma}(a,b)] and the group velocities [see Figs.~\ref{fig: wp sigma}(c,d)].

\red{The results presented} in Fig.~\ref{fig: wp sigma} show that the semi-analytic approach using Eq.~(\ref{eq: finite sigma}) gives good approximations for the wave-packet dynamics. More importantly, it clearly shows that the anomalous drift is strongly modified by the size of the wave packet. Hence, experimentally, the best approach is to start with a tightly focused wave packet in momentum space to minimize finite-size effects.

\section{Conclusion}
In conclusion, we presented two different generalizations of the quantum geometric tensor in non-Hermitian systems and discussed methods for experimentally constructing them in a two-band system, e.g. a photonic or an exciton-polariton system.

In particular, we proposed a method for measuring the left eigenstates from the directly accessible right eigenstates, which opens possibilities to fully explore biorthogonal quantum mechanics. This includes the LR QGT, the Berry curvature, and the Berry connection.

We derived the analytic forms of the components of the RR and LR QGTs of the exciton-polariton model and showed that they are distinct from each other, for example, the RR Berry curvature is non-zero even without the Zeeman splitting while the LR Berry curvature is not. This shows that unlike in the Hermitian case, it is possible to observe anomalous Hall drift in the non-Hermitian exciton-polariton system without an external magnetic field that generates the Zeeman splitting.

We extended the semi-classical equations of motion developed to include both self-acceleration and anomalous Hall drift to investigate how the self-acceleration resulting from the finite wave-packet size interplays with the non-Hermitian anomalous Hall drift. Our results highlight that both LR and RR Berry connection and Berry curvature play direct roles in the anomalous drift.

Our results further show that both the LR and RR QGT components and related quantities, like the Berry connection, play a role in non-Hermitian dynamics. Here, we proposed a method for verifying this role in experiments.

Our work paves the way for probing unanswered questions regarding the non-Hermitian QGT and its components. For example, previous works \cite{bleu2018WP,gao2015WP} showed that the quantum metric appears in the wave-packet dynamics though non-adiabatic corrections or inter-band contribution. It is natural to extend this approach to non-Hermitian systems, to explore the dynamics close to the exceptional points.

\begin{backmatter}

\bmsection{Funding}
Australian Research Council (CE170100039, DE220100712)

\bmsection{Acknowledgments} 
We thank Olivier Bleu for the fruitful discussions. This work was supported by the Australian Research Council (ARC) through the Centre of Excellence Grant CE170100039 and the Discovery Early Career Researcher Award DE220100712. Y.-M. R. Hu is supported by the Australian Government Research Training Program (RTP) Scholarship.

\bmsection{Disclosures}
The authors declare no conflict of interest.

\bmsection{Data availability} Data underlying the results presented in this paper are not publicly available at this time but may be obtained from the authors upon reasonable request.

\bmsection{Supplemental document}
See Supplement 1 for supporting content.

\end{backmatter}

\bibliography{refs}

\begin{thebibliography}{10}
\newcommand{\enquote}[1]{``#1''}

\bibitem{kane2005}
C.~L. Kane and E.~J. Mele, \enquote{${Z}_{2}$ topological order and the quantum
  spin {H}all effect,} {\protect\JournalTitle{Phys. Rev. Lett.}} \textbf{95},
  146802 (2005).

\bibitem{kane2005graphene}
C.~L. Kane and E.~J. Mele, \enquote{Quantum spin {H}all effect in graphene,}
  {\protect\JournalTitle{Phys. Rev. Lett.}} \textbf{95}, 226801 (2005).

\bibitem{xiao2010}
D.~Xiao, M.-C. Chang, and Q.~Niu, \enquote{Berry phase effects on electronic
  properties,} {\protect\JournalTitle{Rev. Mod. Phys.}} \textbf{82}, 1959--2007
  (2010).

\bibitem{senthil2015}
T.~Senthil, \enquote{Symmetry-protected topological phases of quantum matter,}
  {\protect\JournalTitle{Annual Review of Condensed Matter Physics}}
  \textbf{6}, 299--324 (2015).

\bibitem{berry1984}
M.~V. Berry, \enquote{Quantal phase factors accompanying adiabatic changes,}
  {\protect\JournalTitle{Proc. R. Soc. Lond. A}} pp. 45--57 (1984).

\bibitem{chang1995}
M.-C. Chang and Q.~Niu, \enquote{{B}erry phase, hyperorbits, and the
  {H}ofstadter spectrum: Semiclassical dynamics in magnetic {B}loch bands,}
  {\protect\JournalTitle{Phys. Rev. B}} \textbf{53}, 7010--7023 (1996).

\bibitem{sundaram1999}
G.~Sundaram and Q.~Niu, \enquote{Wave-packet dynamics in slowly perturbed
  crystals: Gradient corrections and {B}erry-phase effects,}
  {\protect\JournalTitle{Phys. Rev. B}} \textbf{59}, 14915--14925 (1999).

\bibitem{chang2008}
M.-C. Chang and Q.~Niu, \enquote{Berry curvature, orbital moment, and effective
  quantum theory of electrons in electromagnetic fields,}
  {\protect\JournalTitle{Journal of Physics: Condensed Matter}} \textbf{20},
  193202 (2008).

\bibitem{culcer2005}
D.~Culcer, Y.~Yao, and Q.~Niu, \enquote{Coherent wave-packet evolution in
  coupled bands,} {\protect\JournalTitle{Phys. Rev. B}} \textbf{72}, 085110
  (2005).

\bibitem{gao2014}
Y.~Gao, S.~A. Yang, and Q.~Niu, \enquote{Field induced positional shift of
  {B}loch electrons and its dynamical implications,}
  {\protect\JournalTitle{Phys. Rev. Lett.}} \textbf{112}, 166601 (2014).

\bibitem{gao2015WP}
Y.~Gao, S.~A. Yang, and Q.~Niu, \enquote{Geometrical effects in orbital
  magnetic susceptibility,} {\protect\JournalTitle{Phys. Rev. B}} \textbf{91},
  214405 (2015).

\bibitem{bleu2018WP}
O.~Bleu, G.~Malpuech, Y.~Gao, and D.~D. Solnyshkov, \enquote{Effective theory
  of nonadiabatic quantum evolution based on the quantum geometric tensor,}
  {\protect\JournalTitle{Phys. Rev. Lett.}} \textbf{121}, 020401 (2018).

\bibitem{gianfrate2020}
A.~Gianfrate, O.~Bleu, L.~Dominici, V.~Ardizzone, M.~De~Giorgi, D.~Ballarini,
  G.~Lerario, K.~W. West, L.~N. Pfeiffer, D.~D. Solnyshkov, D.~Sanvitto, and
  G.~Malpuech, \enquote{Measurement of the quantum geometric tensor and of the
  anomalous {H}all drift,} {\protect\JournalTitle{Nature}} \textbf{578},
  381--385 (2020).

\bibitem{leblanc2021}
C.~Leblanc, G.~Malpuech, and D.~D. Solnyshkov, \enquote{Universal semiclassical
  equations based on the quantum metric for a two-band system,}
  {\protect\JournalTitle{Phys. Rev. B}} \textbf{104}, 134312 (2021).

\bibitem{shen2018}
H.~Shen, B.~Zhen, and L.~Fu, \enquote{Topological band theory for
  non-{H}ermitian {H}amiltonians,} {\protect\JournalTitle{Phys. Rev. Lett.}}
  \textbf{120}, 146402 (2018).

\bibitem{bleu2018}
O.~Bleu, D.~D. Solnyshkov, and G.~Malpuech, \enquote{Measuring the quantum
  geometric tensor in two-dimensional photonic and exciton-polariton systems,}
  {\protect\JournalTitle{Phys. Rev. B}} \textbf{97}, 195422 (2018).

\bibitem{solnyshkov2021}
D.~D. Solnyshkov, C.~Leblanc, L.~Bessonart, A.~Nalitov, J.~Ren, Q.~Liao, F.~Li,
  and G.~Malpuech, \enquote{Quantum metric and wave packets at exceptional
  points in non-{H}ermitian systems,} {\protect\JournalTitle{Phys. Rev. B}}
  \textbf{103}, 125302 (2021).

\bibitem{provost1980}
J.~Provost and G.~Vallee, \enquote{Riemannian structure on manifolds of quantum
  states,} {\protect\JournalTitle{Commun.Math. Phys.}} \textbf{76}, 289--301
  (1980).

\bibitem{liao2021}
Q.~Liao, C.~Leblanc, J.~Ren, F.~Li, Y.~Li, D.~Solnyshkov, G.~Malpuech, J.~Yao,
  and H.~Fu, \enquote{Experimental measurement of the divergent quantum metric
  of an exceptional point,} {\protect\JournalTitle{Phys. Rev. Lett.}}
  \textbf{127}, 107402 (2021).

\bibitem{zhang2019}
D.-J. Zhang, Q.-h. Wang, and J.~Gong, \enquote{Quantum geometric tensor in
  $\mathcal{PT}$-symmetric quantum mechanics,} {\protect\JournalTitle{Phys.
  Rev. A}} \textbf{99}, 042104 (2019).

\bibitem{zhu2021}
Y.-Q. Zhu, W.~Zheng, S.-L. Zhu, and G.~Palumbo, \enquote{Band topology of
  pseudo-{H}ermitian phases through tensor {B}erry connections and quantum
  metric,} {\protect\JournalTitle{Phys. Rev. B}} \textbf{104}, 205103 (2021).

\bibitem{gu2010}
S.-J. Gu, \enquote{Fidelity approach to quantum phase transitions,}
  {\protect\JournalTitle{International Journal of Modern Physics B}}
  \textbf{24}, 4371--4458 (2010).

\bibitem{jozsa1994}
R.~Jozsa, \enquote{Fidelity for mixed quantum states,}
  {\protect\JournalTitle{Journal of Modern Optics}} \textbf{41}, 2315--2323
  (1994).

\bibitem{tzeng2021}
Y.-C. Tzeng, C.-Y. Ju, G.-Y. Chen, and W.-M. Huang, \enquote{Hunting for the
  non-{H}ermitian exceptional points with fidelity susceptibility,}
  {\protect\JournalTitle{Phys. Rev. Res.}} \textbf{3}, 013015 (2021).

\bibitem{jiang2018}
H.~Jiang, C.~Yang, and S.~Chen, \enquote{Topological invariants and phase
  diagrams for one-dimensional two-band non-{H}ermitian systems without chiral
  symmetry,} {\protect\JournalTitle{Phys. Rev. A}} \textbf{98}, 052116 (2018).

\bibitem{sun2021}
G.~Sun, J.-C. Tang, and S.-P. Kou, \enquote{Biorthogonal quantum criticality in
  non-{H}ermitian many-body systems,} {\protect\JournalTitle{Frontiers of
  Physics}} \textbf{17}, 33502 (2021).

\bibitem{matsumoto2020}
N.~Matsumoto, K.~Kawabata, Y.~Ashida, S.~Furukawa, and M.~Ueda,
  \enquote{Continuous phase transition without gap closing in non-{H}ermitian
  quantum many-body systems,} {\protect\JournalTitle{Phys. Rev. Lett.}}
  \textbf{125}, 260601 (2020).

\bibitem{tu2023}
Y.-T. Tu, I.~Jang, P.-Y. Chang, and Y.-C. Tzeng, \enquote{General properties of
  fidelity in non-{H}ermitian quantum systems with {PT} symmetry,}
  {\protect\JournalTitle{Quantum}} \textbf{7}, 960 (2023).

\bibitem{ye2023}
C.~C. Ye, W.~L. Vleeshouwers, S.~Heatley, V.~Gritsev, and C.~M. Smith,
  \enquote{Quantum geometry of non-{H}ermitian topological systems,}
  {\protect\JournalTitle{arXiv preprint arXiv:2305.17675}}  (2023).

\bibitem{hetenyi2023}
B.~Het\'enyi and P.~L\'evay, \enquote{Fluctuations, uncertainty relations, and
  the geometry of quantum state manifolds,} {\protect\JournalTitle{Phys. Rev.
  A}} \textbf{108}, 032218 (2023).

\bibitem{el-ganainy2018}
R.~El-Ganainy, K.~G. Makris, M.~Khajavikhan, Z.~H. Musslimani, S.~Rotter, and
  D.~N. Christodoulides, \enquote{Non-{H}ermitian physics and {PT} symmetry,}
  {\protect\JournalTitle{Nature Physics}} \textbf{14}, 11--19 (2018).

\bibitem{ghatak2019}
A.~Ghatak and T.~Das, \enquote{New topological invariants in non-{H}ermitian
  systems,} {\protect\JournalTitle{Journal of Physics: Condensed Matter}}
  \textbf{31}, 263001 (2019).

\bibitem{bergoltz2021}
E.~J. Bergholtz, J.~C. Budich, and F.~K. Kunst, \enquote{{Exceptional topology
  of non-Hermitian systems},} {\protect\JournalTitle{Rev. Mod. Phys.}}
  \textbf{93}, 015005 (2021).

\bibitem{ozdemir2019}
{\c S}.~K. {\"O}zdemir, S.~Rotter, F.~Nori, and L.~Yang, \enquote{Parity--time
  symmetry and exceptional points in photonics,} {\protect\JournalTitle{Nature
  Materials}} \textbf{18}, 783--798 (2019).

\bibitem{kunst2018}
F.~K. Kunst, E.~Edvardsson, J.~C. Budich, and E.~J. Bergholtz,
  \enquote{Biorthogonal bulk-boundary correspondence in non-{H}ermitian
  systems,} {\protect\JournalTitle{Phys. Rev. Lett.}} \textbf{121}, 026808
  (2018).

\bibitem{leykam2017}
D.~Leykam, K.~Y. Bliokh, C.~Huang, Y.~D. Chong, and F.~Nori, \enquote{Edge
  modes, degeneracies, and topological numbers in non-{H}ermitian systems,}
  {\protect\JournalTitle{Phys. Rev. Lett.}} \textbf{118}, 040401 (2017).

\bibitem{gao2015}
T.~Gao, E.~Estrecho, K.~Y. Bliokh, T.~C.~H. Liew, M.~D. Fraser, S.~Brodbeck,
  M.~Kamp, C.~Schneider, S.~H{\"o}fling, Y.~Yamamoto, F.~Nori, Y.~S. Kivshar,
  A.~G. Truscott, R.~G. Dall, and E.~A. Ostrovskaya, \enquote{Observation of
  non-{H}ermitian degeneracies in a chaotic exciton-polariton billiard,}
  {\protect\JournalTitle{Nature}} \textbf{526}, 554--558 (2015).

\bibitem{krol2022}
M.~Kr{\'o}l, I.~Septembre, P.~Oliwa, M.~K{\k e}dziora, K.~{\L}empicka-Mirek,
  M.~Muszy{\'n}ski, R.~Mazur, P.~Morawiak, W.~Piecek, P.~Kula, W.~Bardyszewski,
  P.~G. Lagoudakis, D.~D. Solnyshkov, G.~Malpuech, B.~Pi{\k e}tka, and
  J.~Szczytko, \enquote{Annihilation of exceptional points from different dirac
  valleys in a 2d photonic system,} {\protect\JournalTitle{Nature
  Communications}} \textbf{13}, 5340 (2022).

\bibitem{zhou2018}
H.~Zhou, C.~Peng, Y.~Yoon, C.~W. Hsu, K.~A. Nelson, L.~Fu, J.~D. Joannopoulos,
  M.~Soljačić, and B.~Zhen, \enquote{Observation of bulk {F}ermi arc and
  polarization half charge from paired exceptional points,}
  {\protect\JournalTitle{Science}} \textbf{359}, 1009--1012 (2018).

\bibitem{su2021}
R.~Su, E.~Estrecho, D.~Biega{\'n}ska, Y.~Huang, M.~Wurdack, M.~Pieczarka, A.~G.
  Truscott, T.~C.~H. Liew, E.~A. Ostrovskaya, and Q.~Xiong, \enquote{Direct
  measurement of a non-{H}ermitian topological invariant in a hybrid
  light-matter system,} {\protect\JournalTitle{Science Advances}} \textbf{7},
  eabj8905 (2021).

\bibitem{zhang2020}
K.~Zhang, Z.~Yang, and C.~Fang, \enquote{Correspondence between winding numbers
  and skin modes in non-{H}ermitian systems,} {\protect\JournalTitle{Phys. Rev.
  Lett.}} \textbf{125}, 126402 (2020).

\bibitem{gong2018}
Z.~Gong, Y.~Ashida, K.~Kawabata, K.~Takasan, S.~Higashikawa, and M.~Ueda,
  \enquote{Topological phases of non-{H}ermitian systems,}
  {\protect\JournalTitle{Phys. Rev. X}} \textbf{8}, 031079 (2018).

\bibitem{kawabata2018}
K.~Kawabata, K.~Shiozaki, and M.~Ueda, \enquote{Anomalous helical edge states
  in a non-{H}ermitian {C}hern insulator,} {\protect\JournalTitle{Phys. Rev.
  B}} \textbf{98}, 165148 (2018).

\bibitem{lee2019}
C.~H. Lee and R.~Thomale, \enquote{Anatomy of skin modes and topology in
  non-{H}ermitian systems,} {\protect\JournalTitle{Phys. Rev. B}} \textbf{99},
  201103 (2019).

\bibitem{hofmann2020}
T.~Hofmann, T.~Helbig, F.~Schindler, N.~Salgo, M.~Brzezi\ifmmode~\acute{n}\else
  \'{n}\fi{}ska, M.~Greiter, T.~Kiessling, D.~Wolf, A.~Vollhardt,
  A.~Kaba\ifmmode~\check{s}\else \v{s}\fi{}i, C.~H. Lee, A.~Bilu\ifmmode
  \check{s}\else \v{s}\fi{}i\ifmmode~\acute{c}\else \'{c}\fi{}, R.~Thomale, and
  T.~Neupert, \enquote{Reciprocal skin effect and its realization in a
  topolectrical circuit,} {\protect\JournalTitle{Phys. Rev. Research}}
  \textbf{2}, 023265 (2020).

\bibitem{yao2018}
S.~Yao and Z.~Wang, \enquote{Edge states and topological invariants of
  non-{H}ermitian systems,} {\protect\JournalTitle{Phys. Rev. Lett.}}
  \textbf{121}, 086803 (2018).

\bibitem{zhang2022}
K.~Zhang, Z.~Yang, and C.~Fang, \enquote{Universal non-{H}ermitian skin effect
  in two and higher dimensions,} {\protect\JournalTitle{Nature Communications}}
  \textbf{13}, 2496 (2022).

\bibitem{weidemann2020}
S.~Weidemann, M.~Kremer, T.~Helbig, T.~Hofmann, A.~Stegmaier, M.~Greiter,
  R.~Thomale, and A.~Szameit, \enquote{Topological funneling of light,}
  {\protect\JournalTitle{Science}} \textbf{368}, 311 -- 314 (2020).

\bibitem{fan2020}
G.~H. A.~Fan and S.~Liang, \enquote{Complex {B}erry curvature pair and quantum
  {H}all admittance in non-{H}ermitian systems,} {\protect\JournalTitle{Journal
  of Physics Communications}} \textbf{4}, 115006 (2020).

\bibitem{brody2013}
D.~C. Brody and E.-M. Graefe, \enquote{Information geometry of complex
  {H}amiltonians and exceptional points,} {\protect\JournalTitle{Entropy}}
  \textbf{15}, 3361--3378 (2013).

\bibitem{cuerda2023}
J.~{Cuerda}, J.~M. {Taskinen}, N.~{K{\"a}llman}, L.~{Grabitz}, and
  P.~{T{\"o}rm{\"a}}, \enquote{{Observation of Quantum metric and non-Hermitian
  Berry curvature in a plasmonic lattice},} {\protect\JournalTitle{arXiv
  e-prints}} arXiv:2305.13174 (2023).

\bibitem{xu2017}
Y.~Xu, S.-T. Wang, and L.-M. Duan, \enquote{Weyl exceptional rings in a
  three-dimensional dissipative cold atomic gas,} {\protect\JournalTitle{Phys.
  Rev. Lett.}} \textbf{118}, 045701 (2017).

\bibitem{silberstein2020}
N.~Silberstein, J.~Behrends, M.~Goldstein, and R.~Ilan, \enquote{Berry
  connection induced anomalous wave-packet dynamics in non-{H}ermitian
  systems,} {\protect\JournalTitle{Phys. Rev. B}} \textbf{102}, 245147 (2020).

\bibitem{wang2022}
J.-H. Wang, Y.-L. Tao, and Y.~Xu, \enquote{Anomalous transport induced by
  non-{H}ermitian anomalous {B}erry connection in non-{H}ermitian systems,}
  {\protect\JournalTitle{Chin. Phys. Lett.}} \textbf{39}, 010301 (2022).

\bibitem{tercas2014}
H.~Ter\ifmmode~\mbox{\c{c}}\else \c{c}\fi{}as, H.~Flayac, D.~D. Solnyshkov, and
  G.~Malpuech, \enquote{Non-{A}belian gauge fields in photonic cavities and
  photonic superfluids,} {\protect\JournalTitle{Phys. Rev. Lett.}}
  \textbf{112}, 066402 (2014).

\bibitem{kasprzak2006}
J.~Kasprzak, M.~Richard, S.~Kundermann, A.~Baas, P.~Jeambrun, J.~M.~J. Keeling,
  F.~M. Marchetti, M.~H. Szyma{\'n}ska, R.~Andr{\'e}, J.~L. Staehli, V.~Savona,
  P.~B. Littlewood, B.~Deveaud, and L.~S. Dang, \enquote{{B}ose--{E}instein
  condensation of exciton polaritons,} {\protect\JournalTitle{Nature}}
  \textbf{443}, 409--414 (2006).

\bibitem{deng2010}
H.~Deng, H.~Haug, and Y.~Yamamoto, \enquote{{Exciton-polariton Bose-Einstein
  condensation},} {\protect\JournalTitle{Rev. Mod. Phys.}} \textbf{82},
  1489--1537 (2010).

\bibitem{carusotto2013}
I.~Carusotto and C.~Ciuti, \enquote{Quantum fluids of light,}
  {\protect\JournalTitle{Rev. Mod. Phys.}} \textbf{85}, 299--366 (2013).

\bibitem{klembt2018}
S.~Klembt, T.~H. Harder, O.~A. Egorov, K.~Winkler, R.~Ge, M.~A. Bandres,
  M.~Emmerling, L.~Worschech, T.~C.~H. Liew, M.~Segev, C.~Schneider, and
  S.~H{\"o}fling, \enquote{Exciton-polariton topological insulator,}
  {\protect\JournalTitle{Nature}} \textbf{562}, 552--556 (2018).

\bibitem{pieczarka2021}
M.~Pieczarka, E.~Estrecho, S.~Ghosh, M.~Wurdack, M.~Steger, D.~W. Snoke,
  K.~West, L.~N. Pfeiffer, T.~C.~H. Liew, A.~G. Truscott, and E.~A.
  Ostrovskaya, \enquote{Topological phase transition in an all-optical
  exciton-polariton lattice,} {\protect\JournalTitle{Optica}} \textbf{8},
  1084--1091 (2021).

\bibitem{ren2021}
J.~Ren, Q.~Liao, F.~Li, Y.~Li, O.~Bleu, G.~Malpuech, J.~Yao, H.~Fu, and
  D.~Solnyshkov, \enquote{Nontrivial band geometry in an optically active
  system,} {\protect\JournalTitle{Nature Communications}} \textbf{12}, 689
  (2021).

\bibitem{pickup2020}
L.~{Pickup}, H.~{Sigurdsson}, J.~{Ruostekoski}, and P.~G. {Lagoudakis},
  \enquote{{Synthetic band-structure engineering in polariton crystals with
  non-Hermitian topological phases},} {\protect\JournalTitle{Nature
  Communications}} \textbf{11}, 4431 (2020).

\bibitem{comaron2020}
P.~{Comaron}, V.~{Shahnazaryan}, W.~{Brzezicki}, T.~{Hyart}, and
  M.~{Matuszewski}, \enquote{{Non-Hermitian topological end-mode lasing in
  polariton systems},} {\protect\JournalTitle{Physical Review Research}}
  \textbf{2}, 022051 (2020).

\bibitem{pernet2022}
N.~{Pernet}, P.~{St-Jean}, D.~D. {Solnyshkov}, G.~{Malpuech}, N.~{Carlon
  Zambon}, Q.~{Fontaine}, B.~{Real}, O.~{Jamadi}, A.~{Lema{\^\i}tre},
  M.~{Morassi}, L.~{Le Gratiet}, T.~{Baptiste}, A.~{Harouri}, I.~{Sagnes},
  A.~{Amo}, S.~{Ravets}, and J.~{Bloch}, \enquote{{Gap solitons in a
  one-dimensional driven-dissipative topological lattice},}
  {\protect\JournalTitle{Nature Physics}} \textbf{18}, 678--684 (2022).

\bibitem{sigurdsson2017}
H.~Sigurdsson, G.~Li, and T.~C.~H. Liew, \enquote{Spontaneous and superfluid
  chiral edge states in exciton-polariton condensates,}
  {\protect\JournalTitle{Phys. Rev. B}} \textbf{96}, 115453 (2017).

\bibitem{mandal2020}
S.~{Mandal}, R.~{Banerjee}, E.~A. {Ostrovskaya}, and T.~C.~H. {Liew},
  \enquote{{Nonreciprocal Transport of Exciton Polaritons in a Non-Hermitian
  Chain},} {\protect\JournalTitle{Phys. Rev. Lett.}} \textbf{125}, 123902
  (2020).

\bibitem{xu2021}
X.~{Xu}, H.~{Xu}, S.~{Mandal}, R.~{Banerjee}, S.~{Ghosh}, and T.~C.~H. {Liew},
  \enquote{{Interaction-induced double-sided skin effect in an
  exciton-polariton system},} {\protect\JournalTitle{Phys. Rev. B}}
  \textbf{103}, 235306 (2021).

\bibitem{xu2021b}
H.~{Xu}, K.~{Dini}, X.~{Xu}, R.~{Banerjee}, S.~{Mandal}, and T.~C.~H. {Liew},
  \enquote{{Nonreciprocal exciton-polariton ring lattices},}
  {\protect\JournalTitle{Phys. Rev. B}} \textbf{104}, 195301 (2021).

\bibitem{mandal2022}
S.~Mandal, R.~Banerjee, and T.~C. Liew, \enquote{From the topological spin-hall
  effect to the non-hermitian skin effect in an elliptical micropillar chain,}
  {\protect\JournalTitle{ACS Photonics}} \textbf{9}, 527--539 (2022).

\bibitem{xu2022}
X.~{Xu}, R.~{Bao}, and T.~C.~H. {Liew}, \enquote{{Non-Hermitian topological
  exciton-polariton corner modes},} {\protect\JournalTitle{Phys. Rev. B}}
  \textbf{106}, L201302 (2022).

\bibitem{kokhanchik2023}
P.~{Kokhanchik}, D.~{Solnyshkov}, and G.~{Malpuech}, \enquote{{Non-Hermitian
  skin effect induced by Rashba-Dresselhaus spin-orbit coupling},}
  {\protect\JournalTitle{arXiv e-prints}} arXiv:2303.08483 (2023).

\bibitem{gao2018}
T.~{Gao}, G.~{Li}, E.~{Estrecho}, T.~C.~H. {Liew}, D.~{Comber-Todd},
  A.~{Nalitov}, M.~{Steger}, K.~{West}, L.~{Pfeiffer}, D.~W. {Snoke}, A.~V.
  {Kavokin}, A.~G. {Truscott}, and E.~A. {Ostrovskaya}, \enquote{{Chiral Modes
  at Exceptional Points in Exciton-Polariton Quantum Fluids},}
  {\protect\JournalTitle{Phys. Rev. Lett.}} \textbf{120}, 065301 (2018).

\bibitem{estrecho2016}
E.~{Estrecho}, T.~{Gao}, S.~{Brodbeck}, M.~{Kamp}, C.~{Schneider},
  S.~{H{\"o}fling}, A.~G. {Truscott}, and E.~A. {Ostrovskaya},
  \enquote{{Visualising Berry phase and diabolical points in a quantum
  exciton-polariton billiard},} {\protect\JournalTitle{Scientific Reports}}
  \textbf{6}, 37653 (2016).

\bibitem{li2022}
Y.~Li, X.~Ma, Z.~Hatzopoulos, P.~G. Savvidis, S.~Schumacher, and T.~Gao,
  \enquote{Switching off a microcavity polariton condensate near the
  exceptional point,} {\protect\JournalTitle{ACS Photonics}} \textbf{9},
  2079--2086 (2022).

\bibitem{hu2022}
Y.-M.~R. Hu, E.~A. Ostrovskaya, and E.~Estrecho, \enquote{Wave-packet dynamics
  in a non-{H}ermitian exciton-polariton system,} {\protect\JournalTitle{Phys.
  Rev. B}} \textbf{108}, 115404 (2023).

\bibitem{piechon2016}
F.~Pi\'echon, A.~Raoux, J.-N. Fuchs, and G.~Montambaux, \enquote{Geometric
  orbital susceptibility: Quantum metric without {B}erry curvature,}
  {\protect\JournalTitle{Phys. Rev. B}} \textbf{94}, 134423 (2016).

\bibitem{polimeno2021tuning}
L.~Polimeno, G.~Lerario, M.~De~Giorgi, L.~De~Marco, L.~Dominici, F.~Todisco,
  A.~Coriolano, V.~Ardizzone, M.~Pugliese, C.~T. Prontera, V.~Maiorano,
  A.~Moliterni, C.~Giannini, V.~Olieric, G.~Gigli, D.~Ballarini, Q.~Xiong,
  A.~Fieramosca, D.~D. Solnyshkov, G.~Malpuech, and D.~Sanvitto,
  \enquote{Tuning of the {B}erry curvature in 2d perovskite polaritons,}
  {\protect\JournalTitle{Nature nanotechnology}} \textbf{16}, 1349--1354
  (2021).

\bibitem{lempicka2022electrically}
K.~{\L}empicka-Mirek, M.~Kr{\'o}l, H.~Sigurdsson, A.~Wincukiewicz, P.~Morawiak,
  R.~Mazur, M.~Muszy{\'n}ski, W.~Piecek, P.~Kula, T.~Stefaniuk, M.~Kamińska,
  L.~D. Marco, P.~G. Lagoudakis, D.~Ballarini, D.~Sanvitto, J.~Szczytko, and
  B.~Piętka, \enquote{Electrically tunable {B}erry curvature and strong
  light-matter coupling in liquid crystal microcavities with {2D} perovskite,}
  {\protect\JournalTitle{Science Advances}} \textbf{8}, eabq7533 (2022).

\bibitem{brody2014}
D.~C. Brody, \enquote{Biorthogonal quantum mechanics,}
  {\protect\JournalTitle{Journal of Physics A: Mathematical and Theoretical}}
  \textbf{47}, 035305 (2013).

\bibitem{gardas2016}
B.~Gardas, S.~Deffner, and A.~Saxena, \enquote{Non-hermitian quantum
  thermodynamics,} {\protect\JournalTitle{Scientific Reports}} \textbf{6},
  23408 (2016).

\bibitem{mostafazadeh2007}
A.~Mostafazadeh, \enquote{Time-dependent pseudo-{H}ermitian {H}amiltonians
  defining a unitary quantum system and uniqueness of the metric operator,}
  {\protect\JournalTitle{Physics Letters B}} \textbf{650}, 208--212 (2007).

\bibitem{ju2019}
C.-Y. Ju, A.~Miranowicz, G.-Y. Chen, and F.~Nori, \enquote{Non-{H}ermitian
  {H}amiltonians and no-go theorems in quantum information,}
  {\protect\JournalTitle{Phys. Rev. A}} \textbf{100}, 062118 (2019).

\bibitem{liu2019}
Z.-Z. Liu, R.~A. Henry, M.~T. Batchelor, and H.-Q. Zhou, \enquote{Some
  ground-state expectation values for the free parafermion {Z(N)} spin chain,}
  {\protect\JournalTitle{Journal of Statistical Mechanics: Theory and
  Experiment}} \textbf{2019}, 124002 (2019).

\bibitem{sternheim1972}
M.~M. Sternheim and J.~F. Walker, \enquote{Non-{H}ermitian {H}amiltonians,
  decaying states, and perturbation theory,} {\protect\JournalTitle{Phys. Rev.
  C}} \textbf{6}, 114--121 (1972).

\bibitem{longhi2022}
S.~Longhi, \enquote{Non-{H}ermitian skin effect and self-acceleration,}
  {\protect\JournalTitle{Phys. Rev. B}} \textbf{105}, 245143 (2022).

\bibitem{yuce2017}
C.~Yuce and Z.~Turker, \enquote{Self-acceleration in non-{H}ermitian systems,}
  {\protect\JournalTitle{Physics Letters A}} \textbf{381}, 2235--2238 (2017).

\bibitem{taha2005}
T.~R. Taha and X.~Xu, \enquote{Parallel split-step {F}ourier methods for the
  coupled nonlinear {S}chr{\"o}dinger type equations,}
  {\protect\JournalTitle{The Journal of Supercomputing}} \textbf{32}, 5--23
  (2005).

\bibitem{weideman1986}
J.~A. Weideman and B.~M. Herbst, \enquote{Split-step methods for the solution
  of the nonlinear {S}chr{\"o}dinger equation,} {\protect\JournalTitle{SIAM
  Journal on Numerical Analysis}} \textbf{23}, 485--507 (1986).

\bibitem{sedov2018}
E.~S. Sedov, Y.~G. Rubo, and A.~V. Kavokin, \enquote{$\mathit{Zitterbewegung}$
  of exciton-polaritons,} {\protect\JournalTitle{Phys. Rev. B}} \textbf{97},
  245312 (2018).

\bibitem{pan2020}
L.~Pan, X.~Chen, Y.~Chen, and H.~Zhai, \enquote{Non-{H}ermitian linear response
  theory,} {\protect\JournalTitle{Nature Physics}} \textbf{16}, 767--771
  (2020).

\end{thebibliography}


\begin{thebibliography}{10}
\newcommand{\enquote}[1]{``#1''}

\bibitem{matsumoto2020}
N.~Matsumoto, K.~Kawabata, Y.~Ashida, S.~Furukawa, and M.~Ueda,
  \enquote{Continuous phase transition without gap closing in non-{H}ermitian
  quantum many-body systems,} {\protect\JournalTitle{Phys. Rev. Lett.}}
  \textbf{125}, 260601 (2020).

\bibitem{sun2021}
G.~Sun, J.-C. Tang, and S.-P. Kou, \enquote{Biorthogonal quantum criticality in
  non-{H}ermitian many-body systems,} {\protect\JournalTitle{Frontiers of
  Physics}} \textbf{17}, 33502 (2021).

\bibitem{tzeng2021}
Y.-C. Tzeng, C.-Y. Ju, G.-Y. Chen, and W.-M. Huang, \enquote{Hunting for the
  non-{H}ermitian exceptional points with fidelity susceptibility,}
  {\protect\JournalTitle{Phys. Rev. Res.}} \textbf{3}, 013015 (2021).

\bibitem{zhang2019}
D.-J. Zhang, Q.-h. Wang, and J.~Gong, \enquote{Quantum geometric tensor in
  $\mathcal{PT}$-symmetric quantum mechanics,} {\protect\JournalTitle{Phys.
  Rev. A}} \textbf{99}, 042104 (2019).

\bibitem{jiang2018}
H.~Jiang, C.~Yang, and S.~Chen, \enquote{Topological invariants and phase
  diagrams for one-dimensional two-band non-{H}ermitian systems without chiral
  symmetry,} {\protect\JournalTitle{Phys. Rev. A}} \textbf{98}, 052116 (2018).

\bibitem{tu2023}
Y.-T. Tu, I.~Jang, P.-Y. Chang, and Y.-C. Tzeng, \enquote{General properties of
  fidelity in non-{H}ermitian quantum systems with {PT} symmetry,}
  {\protect\JournalTitle{Quantum}} \textbf{7}, 960 (2023).

\bibitem{solnyshkov2021}
D.~D. Solnyshkov, C.~Leblanc, L.~Bessonart, A.~Nalitov, J.~Ren, Q.~Liao, F.~Li,
  and G.~Malpuech, \enquote{Quantum metric and wave packets at exceptional
  points in non-{H}ermitian systems,} {\protect\JournalTitle{Phys. Rev. B}}
  \textbf{103}, 125302 (2021).

\bibitem{liao2021}
Q.~Liao, C.~Leblanc, J.~Ren, F.~Li, Y.~Li, D.~Solnyshkov, G.~Malpuech, J.~Yao,
  and H.~Fu, \enquote{Experimental measurement of the divergent quantum metric
  of an exceptional point,} {\protect\JournalTitle{Phys. Rev. Lett.}}
  \textbf{127}, 107402 (2021).

\bibitem{zhu2021}
Y.-Q. Zhu, W.~Zheng, S.-L. Zhu, and G.~Palumbo, \enquote{Band topology of
  pseudo-{H}ermitian phases through tensor {B}erry connections and quantum
  metric,} {\protect\JournalTitle{Phys. Rev. B}} \textbf{104}, 205103 (2021).

\bibitem{lloyd1837}
H.~Lloyd, \enquote{On the phenomena presented by light in its passage along the
  axes of biaxial crystals,} {\protect\JournalTitle{Transactions of the Royal
  Irish Academy}} \textbf{17}, 145--158 (1837).

\bibitem{hamilton1837}
W.~R. Hamilton, \enquote{Third supplement to an essay on systems of rays,}
  {\protect\JournalTitle{Transactions of the Royal Irish Academy}} \textbf{17},
  1--144 (1837).

\bibitem{castro2009}
A.~H. Castro~Neto, F.~Guinea, N.~M.~R. Peres, K.~S. Novoselov, and A.~K. Geim,
  \enquote{The electronic properties of graphene,} {\protect\JournalTitle{Rev.
  Mod. Phys.}} \textbf{81}, 109--162 (2009).

\bibitem{novoselov2004}
K.~S. Novoselov, A.~K. Geim, S.~V. Morozov, D.~Jiang, Y.~Zhang, S.~V. Dubonos,
  I.~V. Grigorieva, and A.~A. Firsov, \enquote{Electric field effect in
  atomically thin carbon films,} {\protect\JournalTitle{Science}} \textbf{306},
  666--669 (2004).

\bibitem{novoselov2005}
K.~S. Novoselov, A.~K. Geim, S.~V. Morozov, D.~Jiang, M.~I. Katsnelson, I.~V.
  Grigorieva, S.~V. Dubonos, and A.~A. Firsov, \enquote{Two-dimensional gas of
  massless {D}irac fermions in graphene,} {\protect\JournalTitle{Nature}}
  \textbf{438}, 197--200 (2005).

\bibitem{su2021}
R.~Su, E.~Estrecho, D.~Biega{\'n}ska, Y.~Huang, M.~Wurdack, M.~Pieczarka, A.~G.
  Truscott, T.~C.~H. Liew, E.~A. Ostrovskaya, and Q.~Xiong, \enquote{Direct
  measurement of a non-{H}ermitian topological invariant in a hybrid
  light-matter system,} {\protect\JournalTitle{Science Advances}} \textbf{7},
  eabj8905 (2021).

\bibitem{gao2015}
T.~Gao, E.~Estrecho, K.~Y. Bliokh, T.~C.~H. Liew, M.~D. Fraser, S.~Brodbeck,
  M.~Kamp, C.~Schneider, S.~H{\"o}fling, Y.~Yamamoto, F.~Nori, Y.~S. Kivshar,
  A.~G. Truscott, R.~G. Dall, and E.~A. Ostrovskaya, \enquote{Observation of
  non-{H}ermitian degeneracies in a chaotic exciton-polariton billiard,}
  {\protect\JournalTitle{Nature}} \textbf{526}, 554--558 (2015).

\bibitem{shen2018}
H.~Shen, B.~Zhen, and L.~Fu, \enquote{Topological band theory for
  non-{H}ermitian {H}amiltonians,} {\protect\JournalTitle{Phys. Rev. Lett.}}
  \textbf{120}, 146402 (2018).

\bibitem{fan2020}
G.~H. A.~Fan and S.~Liang, \enquote{Complex {B}erry curvature pair and quantum
  {H}all admittance in non-{H}ermitian systems,} {\protect\JournalTitle{Journal
  of Physics Communications}} \textbf{4}, 115006 (2020).

\bibitem{hu2022}
Y.-M.~R. Hu, E.~A. Ostrovskaya, and E.~Estrecho, \enquote{Wave-packet dynamics
  in a non-{H}ermitian exciton-polariton system,} {\protect\JournalTitle{Phys.
  Rev. B}} \textbf{108}, 115404 (2023).

\bibitem{silberstein2020}
N.~Silberstein, J.~Behrends, M.~Goldstein, and R.~Ilan, \enquote{Berry
  connection induced anomalous wave-packet dynamics in non-{H}ermitian
  systems,} {\protect\JournalTitle{Phys. Rev. B}} \textbf{102}, 245147 (2020).

\end{thebibliography}

\end{document}


\title{Supplemental Document: \\
Generalized Quantum Geometric Tensor in a Non-Hermitian Exciton-Polariton System}

\author{Y.-M. Robin Hu$^*$, Elena A. Ostrovskaya$^\dagger$, and Eliezer Estrecho$^\ddagger$}

\address{\authormark{1}ARC Centre of Excellence in Future Low-Energy Electronics Technologies and Department of Quantum Science and Technology, Research School of Physics, The Australian National University, Canberra 2601, Australia}

\email{\authormark{*}yow-ming.hu@anu.edu.au, \authormark{$\dagger$}elena.ostrovskaya@anu.edu.au, \authormark{$\ddagger$}eliezer.estrecho@anu.edu.au} 


\section{Different methods of generalizing fidelity in non-Hermitian systems}
Similar to the quantum geometric tensor (QGT), there are also different ways to generalize the \emph{fidelity} and \emph{fidelity susceptibility} to non-Hermitian systems. In Ref.~\cite{matsumoto2020}, the authors defined the fidelity using only the right eigenstates
\begin{equation*}
    \begin{split}
        \mathcal{F}^{RR}(|\psi_n^R(\lambda)\rangle,|\psi_n^R(\lambda+d\lambda)\rangle)&=\langle\psi_n^R(\lambda)|\psi_n^R(\lambda+d\lambda)\rangle\langle\psi_n^R(\lambda+d\lambda)|\psi_n^R(\lambda)\rangle\\
        &=1-g_{n,\mu\nu}^{RR}d\lambda_\mu d\lambda_\nu+...
        \end{split}
\end{equation*}
which we denote as the RR fidelity. In other works, the fidelity was defined using both the left and right eigenstates; in Ref.~\cite{sun2021}, the fidelity is defined as
\begin{equation*}
    \begin{split}
        \mathcal{F}^{LR}(|\psi_n^R(\lambda)\rangle,|\psi_n^R(\lambda+d\lambda)\rangle)&=\sqrt{\langle\psi_n^L(\lambda)|\psi_n^R(\lambda+d\lambda)\rangle\langle\psi_n^L(\lambda+d\lambda)|\psi_n^R(\lambda)\rangle}\\
        &=1-\frac{1}{2}g_{n,\mu\nu}^{LR}d\lambda_\mu d\lambda_\nu+...,
    \end{split}
\end{equation*}
in Ref.~\cite{tzeng2021}, the authors used the definition
\begin{equation*}
    \begin{split}
        \mathcal{F}^{LR}(|\psi_n^R(\lambda)\rangle,|\psi_n^R(\lambda+d\lambda)\rangle)&=\langle\psi_n^L(\lambda)|\psi_n^R(\lambda+d\lambda)\rangle\langle\psi_n^L(\lambda+d\lambda)|\psi_n^R(\lambda)\rangle
        \\
        &=1-g_{n,\mu\nu}^{LR}d\lambda_\mu d\lambda_\nu+...,
    \end{split}
\end{equation*}
in Ref.~\cite{zhang2019}, the authors defined the fidelity as
\begin{equation*}
    \begin{split}
        \mathcal{F}^{LR}(|\psi_n^R(\lambda)\rangle,|\psi_n^R(\lambda+d\lambda)\rangle)&=\sqrt{|\langle\psi_n^L(\lambda)|\psi_n^R(\lambda+d\lambda)\rangle\langle\psi_n^L(\lambda+d\lambda)|\psi_n^R(\lambda)\rangle|}\\
        &=1-\frac{1}{2}\Tilde{g}_{n,\mu\nu}^{LR}d\lambda_\mu d\lambda_\nu+...,
    \end{split}
\end{equation*}
while in Ref.~\cite{jiang2018}, the fidelity is defined as
\begin{equation*}
    \begin{split}
         \mathcal{F}^{LR}(|\psi_n^R(\lambda)\rangle,|\psi_n^R(\lambda+d\lambda)\rangle)&=\frac{1}{2}|\langle\psi_n^L(\lambda)|\psi_n^R(\lambda+d\lambda)\rangle+\langle\psi_n^R(\lambda)|\psi_n^L(\lambda+d\lambda)\rangle|\\
        &=1-\frac{1}{2}\Tilde{g}_{n,\mu\nu}^{LR}d\lambda_\mu d\lambda_\nu+....
    \end{split}
\end{equation*}
Since the fidelity is not the focus of this work, we refer to all of them as the LR fidelity in the main text. For a more detailed discussion on fidelity in non-Hermitian systems, please refer to Ref.~\cite{tu2023}.

\section{Generalized Quantum Geometric Tensor of the Non-Hermitian Dirac Model}
In this supplementary material, we present the two generalized QGT of a non-Hermitian Dirac model, similar to the model used in Ref. \cite{solnyshkov2021}. This model presents a good approximation of the bands near the pair of exceptional points in the exciton-polariton system. As in the main text, we consider two non-Hermitian generalizations of quantum geometric tensor, $Q^{RR}_{\pm,\mu\nu}$, which was used in Refs.~\cite{solnyshkov2021,liao2021}  and $Q^{NH}_{\pm,\mu\nu}$, which was used in Refs.~\cite{zhang2019,zhu2021}
\begin{equation}\tag{S1}
    \begin{split}
        Q^{\alpha\beta}_{\pm,\mu\nu}=&\langle\partial_\mu\psi_\pm^\alpha|\partial_\nu\psi_\pm^\beta\rangle-\langle\partial_\mu\psi_\pm^\alpha|\psi_\pm^\beta\rangle\langle\psi_\pm^\alpha|\partial_\nu\psi_\pm^\beta\rangle
    \end{split}
\end{equation}
where $\alpha,\beta=L,R$ and the eigenstates are normalized as $\langle\psi^\alpha_\pm |\psi^\beta_\pm\rangle=1$ for $Q^{\alpha\beta}_\pm$. We denote $Q^{RR}_{\pm,\mu\nu}$ as the right-right (RR) quantum geometric tensor and $Q^{LR}_{\pm,\mu\nu}$ the left-right (LR) quantum geometric tensor. Similar to the Hermitian limit, the anti-symmetric parts of these non-Hermitian QGTs are related to the non-Hermitian Berry curvature, $i(Q_{\pm,xy}^{\alpha\beta}-Q_{\pm,yx}^{\alpha\beta})=\Omega_\pm^{z,\alpha\beta}$, and we denote their symmetric part as the quantum metric tensor $\frac{1}{2}(Q_{\pm,\mu\nu}^{\alpha\beta}+Q_{\pm,\nu\mu}^{\alpha\beta})=g_{\pm,\mu\nu}^{\alpha\beta}$.

We consider a non-Hermitian generalization of the Dirac model, which has been used to describe the conical diffraction \cite{lloyd1837,hamilton1837} and electronic properties in graphene \cite{castro2009,novoselov2004,novoselov2005}. The system is described by the Hamiltonian:
\begin{equation}\tag{S2}
    \begin{split}
        \mathbf{H}(\mathbf{k})&={\mathbf{h}}(\mathbf{k})\cdot\boldsymbol{\sigma}\\
        {\mathbf{h}}(\mathbf{k})&=[k_x,k_y-i\kappa,\Delta].
    \end{split}
\end{equation}
which has the eigenenergies:
\begin{equation}\tag{S3}
\begin{split}
     E_\pm&=\pm\sqrt{k_x^2+(k_y-i\kappa)^2+\Delta^2}\\
     &=\pm E.
\end{split}
\end{equation}
For $Q^{RR}_{\pm,\mu\nu}$, we require the right eigenstates to be normalized as $\langle\bar{\psi}^R_\pm|\bar{\psi}^R_\pm\rangle=1$, and thus
\begin{equation}\tag{S4}
    \begin{split}
        |\bar{\psi}^R_\pm\rangle=&\frac{1}{\sqrt{(\Delta\pm E)(\Delta\pm E^*)+k_y^2+(k_x+\kappa)^2}}\begin{pmatrix}\Delta\pm E\\ k_x+i(k_y-i\kappa)\end{pmatrix},
    \end{split}
\end{equation}
while for $Q^{LR}_{\pm,\mu\nu}$, we require the eigenstates to be bi-orthonormal $\langle\psi^L_\pm|\psi^R_\pm\rangle=1$, and thus that the eigenstates take the forms as:
\begin{equation}\tag{S5}
    \begin{split}
        |\psi^R_\pm\rangle=&\frac{1}{\sqrt{(\Delta\pm E)^2+k_x^2+(k_y-i\kappa)^2}}\begin{pmatrix}\Delta\pm E\\ k_x+i(k_y-i\kappa)\end{pmatrix}\\
        |\psi^L_\pm\rangle=&\frac{1}{\sqrt{(\Delta\pm E^*)^2+k_x^2+(k_y+i\kappa)^2}}\begin{pmatrix}\Delta\pm E^*\\ k_x+i(k_y+i\kappa)\end{pmatrix}.
    \end{split}
\end{equation}

When $\Delta=0$, the term $-i\kappa\sigma_y$ splits the Dirac point at the origin into a pair of non-Hermitian degeneracies known as exceptional points \cite{su2021,gao2015} at $\mathbf{k}=(\pm\kappa,0)$. When $|\Delta|>0$ is included, it will shrink the bulk Fermi arc and move the exceptional points to $\mathbf{k}=(\pm\sqrt{\kappa^2-\Delta^2},0)$. The exceptional points will merge to a hybrid point \cite{shen2018} at $|\Delta|=\kappa$ and will be destroyed when $|\Delta|>\kappa$.

The components of the RR QGT take the forms 
\begin{equation}\tag{S6}
    \begin{split}
        g^{RR}_{\pm,xx}&=\frac{|X_\pm|^2}{|E|^4\left((\Delta\pm E)^2+k_y^2+(k_x+\kappa)^2\right)^2}\\
        g^{RR}_{\pm,xy}&=\frac{\operatorname{Im}X_\pm\operatorname{Re}Y_\pm-\operatorname{Im}Y_\pm\operatorname{Re}X_\pm}{|E|^4\left((\Delta\pm E)^2+k_y^2+(k_x+\kappa)^2\right)^2}\\
        g^{RR}_{\pm,yy}&=\frac{|Y_\pm|^2}{|E|^4\left((\Delta\pm E)^2+k_y^2+(k_x+\kappa)^2\right)^2}\\
        \Omega^{z,RR}_{\pm,xx}&=-2\frac{\operatorname{Re}X_\pm\operatorname{Re}Y_\pm+\operatorname{Im}X_\pm\operatorname{Im}Y_\pm}{|E|^4\left((\Delta\pm E)^2+k_y^2+(k_x+\kappa)^2\right)^2}\\
                X_\pm&=\mp k_x E(ik_y+\kappa)+k_x^2\Delta+(\Delta\pm E)(\Delta^2+(k_y-i\kappa))\\
        Y_\pm&=\pm k_x E(k_x+i k_y+\kappa)+\Delta(k_x^2+\Delta(\Delta\pm E)+(k_y-i\kappa)^2),
    \end{split}
\end{equation}
where $X_\pm$, $Y_\pm$ are used to simplify the expressions.

When $\Delta=0$, the components of the RR QGT of the non-Hermitian Dirac model can be simplified as
\begin{equation}\tag{S7}
    \begin{split}
        g^{RR}_{\pm,xx}&=\frac{k_y^2+\kappa^2}{E E^*\Big(\sqrt{k_y^2+(k_x+\kappa)^2}+\sqrt{k_y^2+(k_x-\kappa)^2}\Big)^2}\\
        g^{RR}_{\pm,xy}&=-\frac{k_x k_y}{E E^*\Big(\sqrt{k_y^2+(k_x+\kappa)^2}+\sqrt{k_y^2+(k_x-\kappa)^2}\Big)^2}\\
        g^{RR}_{\pm,yy}&=\frac{k_x^2}{E E^*\Big(\sqrt{k_y^2+(k_x+\kappa)^2}+\sqrt{k_y^2+(k_x-\kappa)^2}\Big)^2}\\
        \Omega^{z,RR}_{\pm,xx}&=\frac{2 k_x \kappa}{E E^*\Big(\sqrt{k_y^2+(k_x+\kappa)^2}+\sqrt{k_y^2+(k_x-\kappa)^2}\Big)^2}.
    \end{split}
\end{equation}
Similar to the results obtained using the exciton-polariton model in the main text, all components of the RR QGT of the two bands agree at $\Delta=0$. Furthermore, we can observe that when $\Delta=0$, all components of the RR QGT of the non-Hermitian Dirac model diverge as $k^{-1}$ at each exceptional point. This is similar to the behavior of the RR QGT of the exciton polaritons in the main text and agrees with the results in Ref~\cite{solnyshkov2021}.

The non-Hermitian quantum geometric tensor $Q^{LR}_{\pm,\mu\nu}$ and Berry curvature $\Omega_\pm^{z,LR}$ of this non-Hermitian Dirac model take the forms
\begin{equation}\tag{S8}
    \begin{split}
        g^{LR}_{\pm,xx}=&\frac{(k_y-i\kappa)^2+\Delta^2}{4E^4}\\
        g^{LR}_{\pm,xy}=&\frac{k_x(k_y-i\kappa)}{4E^4}\\
        g^{LR}_{\pm,yy}=&\frac{kx^2+\Delta^2}{4E^4}\\
        \Omega^{z,LR}_{\pm}=&\mp\frac{\Delta}{2E^3}
    \end{split}
\end{equation}
Unlike in $Q^{RR}_{\pm,\mu\nu}$, $g^{LR}_{\pm,\mu\nu}$ diverge at each exceptional point as $k^{-2}$, while $\Omega^{z,LR}_\pm$ diverge as $k^{-3/2}$, which is similar to the results obtained using the exciton-polariton model in the main text. We also note that $\Omega_\pm^{z,LR}$ of the non-Hermitian Dirac model has previously been calculated in Ref.~\cite{fan2020}.

\section{Values of parameters used in the main text}
In this work, we chose the mean exciton-polariton energy at $\mathbf{k}=0$ to be $E_0=2.306$ eV, while the effective polariton mass satisfies $\hbar^2/2m\approx2.3\times10^{-3}$ $\mu$m$^2$eV. 
We chose the parameters corresponding to the splitting of the linearly polarized modes to be $\alpha=8\times 10^{-3}$ eV and $a=10^{-5}$ eV. The parameters for the TE-TM splitting are chosen to be $\beta=10^{-3}$ $\mu$m$^2$eV and $b=7.5\times 10^{-4}$ $\mu$m$^2$eV. 

We expand the exciton-polariton linewidth to the 4th order $\gamma(\mathbf{k})=\gamma_0+\gamma_2\mathbf{k}^4+\gamma_4\mathbf{k}^2$~\cite{hu2022} and choose $\gamma_0=4.5\times10^{-4}$, $\gamma_2=0$ and $\gamma_4=3.75\times 10^{-4}$ $\mu$m$^4$eV. The 4-th order term is necessary to capture the non-monotonic behaviour of the linewidth with increasing $k$ observed in experiments~\cite{su2021}. The mean linewidth should not affect the measurement of the QGT since it is not part of the effective magnetic field. However, it does strongly affect the self-acceleration and hence will interplay with the anomalous Hall effect. If the mean linewidth is constant, the self-acceleration persists (due to the complex effective field) but its direction and magnitude will change.

\section{Wave-packet real and momentum-space distributions}
\begin{figure}[h]
    \centering
    \includegraphics[width=0.9\textwidth]{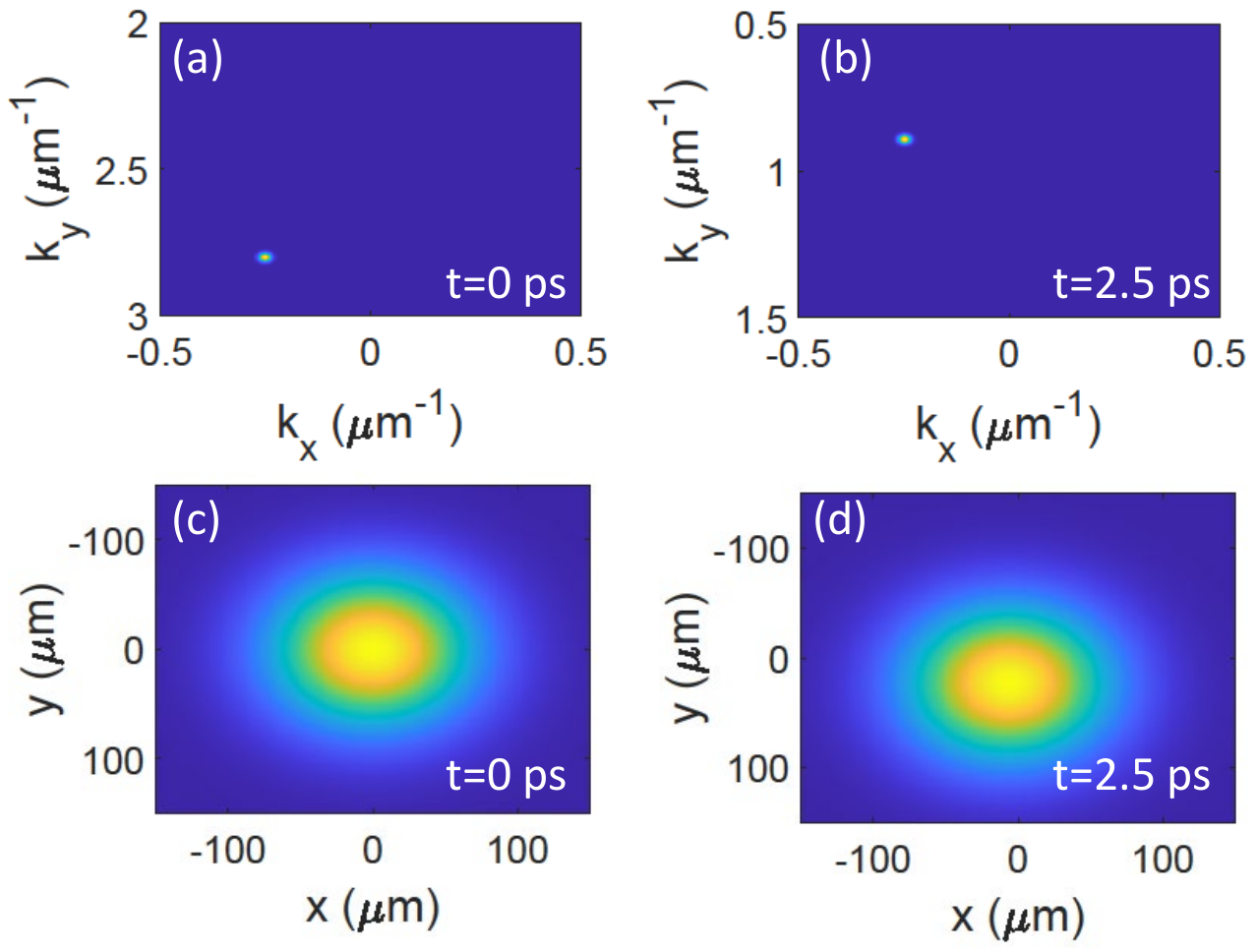}
    \caption{Example distributions of a wave packet at $t=0$ ps and $t=2.5$ ps in the momentum (a,b) and position (c,d) spaces. These correspond to simulation presented in Fig.~8 of the main text.}
    \label{fig: wprk}
\end{figure}

We calculated the time evolution of a Gaussian wavepacket using Eq. 36 of the main text. Sample distributions in position and momentum spaces of the wavepacket are shown in Fig.~\ref{fig: wprk}. Images are shown for the initial ($t=0$) and final distributions after $2.5$ ps time evolution. These images correspond to the trajectories of the wave packet in Fig. 8 of the main text.

\section{Pseudospins for measuring RR and LR quantum geometric tensors}
As discussed in the main text, the pseudospins of the right eignstates are all real-valued while the LR pseudospins can be complex-valued despite the bi-orthogonality condition. In this section, we expand the general forms of $\mathbf{S}^R_\pm$, $\mathbf{S}^{LR}_\pm$ of the general two-band non-Hermitian model presented in the main text.

Recall that the un-normalized right and left eigenstates of the general Hamiltonian $\mathbf{H}=[h_x,h_y,h_z]\cdot[\sigma_x,\sigma_y,\sigma_z]$ take the forms: 
\begin{equation*}\tag{S9}
    |R_\pm\rangle=\begin{pmatrix}h_z\pm \lambda\\h_+\end{pmatrix}
    \qquad \textnormal{and} \qquad
    |L_\pm\rangle=\begin{pmatrix}h_z^*\pm \lambda^*\\h_-^*\end{pmatrix}
\end{equation*}
where $h_\pm=h_x\pm i h_y$.

By expanding $S_{j,\pm}^R=\langle R_\pm|\sigma_j|R_\pm\rangle/\langle R_\pm|R_\pm\rangle$, the components of the pseudospins of the right eignstates take the general forms as:
\begin{equation*}\tag{S10}
    \begin{split}
        S_{x,\pm}^R&=\frac{2\operatorname{Re}[h_+(h_z^*\pm\lambda^*)]}{|h_z\pm\lambda|^2+|h_+|^2}\\
        S_{y,\pm}^R&=\frac{2\operatorname{Im}[h_+(h_z^*\pm\lambda^*)]}{|h_z\pm\lambda|^2+|h_+|^2}\\
        S_{z,\pm}^R&=\frac{|h_z\pm\lambda|^2-|h_+|^2}{|h_z\pm\lambda|^2+|h_+|^2}
    \end{split}
\end{equation*}
which are all real-valued. 

Similarly, by expanding $S_{j,\pm}^{LR}=\langle L_\pm|\sigma_j|R_\pm\rangle/\langle L_\pm|R_\pm\rangle$, the LR pseudospins take the general forms as:
\begin{equation*}\tag{S11}
    \begin{split}
        S_{x,\pm}^{LR}&=\frac{2h_x(h_z\pm\lambda)}{(h_z\pm\lambda)^2+h_x^2+h_y^2}\\
        S_{y,\pm}^{LR}&=\frac{2h_y(h_z\pm\lambda)}{(h_z\pm\lambda)^2+h_x^2+h_y^2}\\
        S_{z,\pm}^{LR}&=\frac{(h_z\pm\lambda)^2-h_x^2-h_y^2}{(h_z\pm\lambda)^2+h_x^2+h_y^2}.
    \end{split}
\end{equation*}
In the Hermitian limit, $\operatorname{Re}h_+=h_x$, while $\operatorname{Im}h_+=h_y$, therefore $\mathbf{S}^{LR}_\pm$ agrees with $\mathbf{S}^R_\pm$. However, in a non-Hermitian system, the effective magnetic fields $h_j$ are generally complex-valued, therefore the components of $\mathbf{S}^{LR}_\pm$ can be complex-valued as well. As a consequence, the angles $\phi^{LR}_\pm$, $\theta^{LR}_\pm$ also become complex valued as shown in the main text.

\section{Measurement of non-Hermitian Berry connection}
The RR and LR Berry connections are defined as
\begin{equation*}\tag{S12}
    \begin{split}
    \mathbf{A}_\pm^{RR}&=\frac{\langle\psi_\pm^R|i\nabla_\mathbf{k}\psi_\pm^R\rangle}{\langle\psi_\pm^R|\psi_\pm^R\rangle}\\
    \mathbf{A}_\pm^{LR}&=\langle\psi_\pm^L|i\nabla_\mathbf{k}\psi_\pm^R\rangle
    \end{split}
\end{equation*}
where $|\psi_\pm^R\rangle$ are normalized with respects to $\langle\psi^L_\pm|$ as $\langle\psi_\pm^L|\psi_\pm^R\rangle=1$ \cite{silberstein2020}. Although neither $\mathbf{A}^{RR}_\pm$ or $\mathbf{A}^{LR}_\pm$ is gauge-invariant by itself, previous work \cite{silberstein2020} has shown that the difference $\mathbf{A}^{RR}_\pm-\mathbf{A}^{LR}_\pm$ is indeed gauge-invariant and is a physical observable. 

Similar to the RR and LR QGTs in the main texts, the RR and LR Berry connections can be constructed using the polarization of photon emission, which can be experimentally measured. The quantity $\mathbf{A}^{RR}_\pm-\mathbf{A}^{LR}_\pm$ can be written in terms of the complex-valued angles of $\mathbf{S}^{LR}_\pm$, namely $\theta_\pm^{LR}$ and $\phi_\pm^{LR}$, as
\begin{equation*}\tag{S13}
    \begin{split}
        \mathbf{A}^{RR}_\pm-\mathbf{A}^{LR}_\pm =& \frac{\cosh{\operatorname{Im}\phi^{LR}_\pm}\sinh{\operatorname{Im}\theta^{LR}_\pm}-i\sin{\operatorname{Re}\theta^{LR}_\pm}\sinh{\operatorname{Im}\phi^{LR}_\pm}}{2\cosh{\operatorname{Im}\theta^{LR}_\pm}\cosh{\operatorname{Im}\phi^{LR}_\pm}+2\cos{\operatorname{Re}\theta^{LR}_\pm}\sinh{\operatorname{Im}\phi^{LR}_\pm}}\nabla_\mathbf{k}\theta^{LR}_\pm\\
        &+\left(\frac{e^{\operatorname{Im}\phi^{LR}_\pm}\left(\cos{\operatorname{Re}\theta^{LR}_\pm}+\cosh{\operatorname{Im}\theta^{LR}_\pm}\right)}{2\cosh{\operatorname{Im}\theta^{LR}_\pm}\cosh{\operatorname{Im}\phi^{LR}_\pm}+2\cos{\operatorname{Re}\theta^{LR}_\pm}\sinh{\operatorname{Im}\phi^{LR}_\pm}}-\cos^2{\frac{\theta^{LR}_\pm}{2}}\right)\nabla_\mathbf{k}\phi^{LR}_\pm.
    \end{split}
\end{equation*}
Since both $\phi^{LR}_\pm$ and $\theta^{LR}_\pm$ can be constructed by measuring the Stokes vector $\mathbf{S}^R_\pm$, this shows that $\mathbf{A}^{RR}_\pm-\mathbf{A}^{LR}_\pm$ can be experimentally measured.

\begin{figure}[h!]
    \centering
    \includegraphics[width=0.9\textwidth]{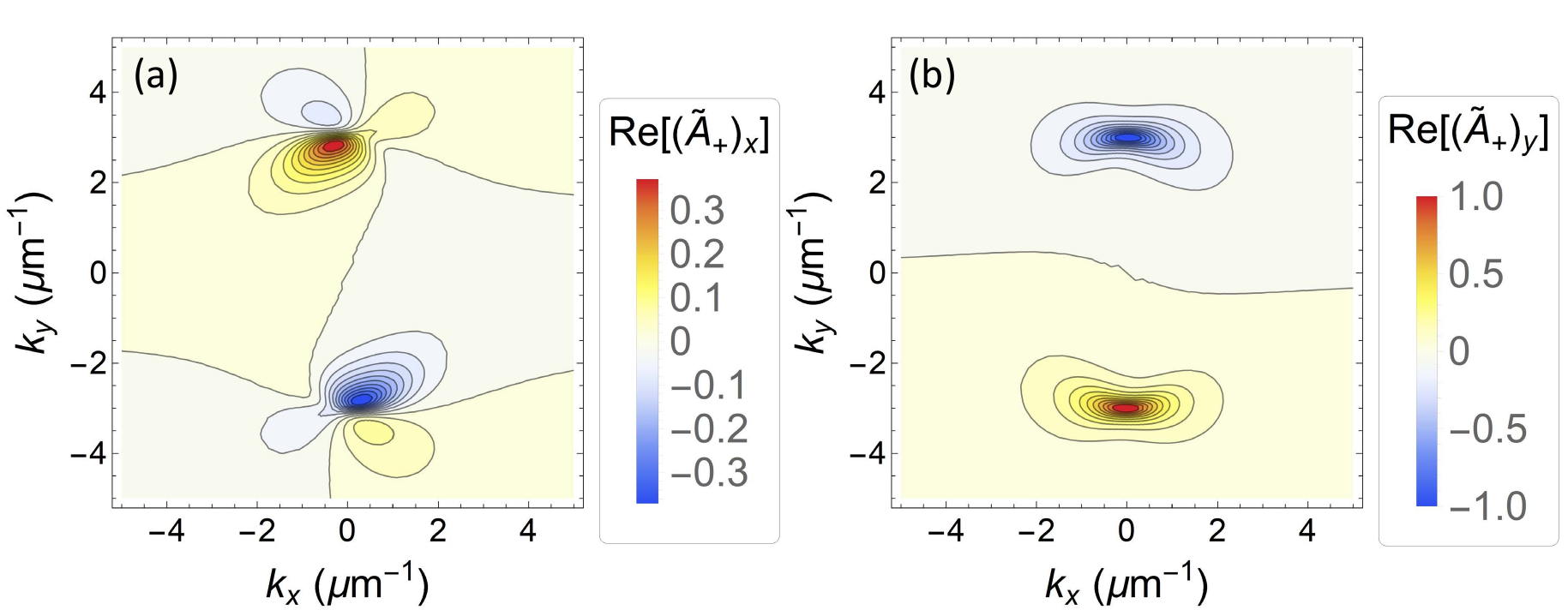}
    \caption{Non-Hermitian Berry connection $\tilde{\mathbf{A}}=\mathbf{A^{RR}-A^{LR}}$ of the upper band of the exciton-polariton system considered in the main text. The $x$ and $y$ components are presented separately.}
    \label{fig: wpa}
\end{figure}

Note that in the main text, we construct the RR QGT with $|\bar{\psi}^{LR}_\pm\rangle$, which satisfy $\langle\bar{\psi}_\pm^R|\bar{\psi}_\pm^R\rangle=1$. However, the RR Berry connection is defined using $|\psi^R_\pm\rangle$, satisfying $\langle\psi_\pm^L|\psi_\pm^R\rangle=1$, instead. To understand why this is necessary, we can construct $|\bar{\psi}^{LR}_\pm\rangle$ from $|\psi^R_\pm\rangle$ as
\begin{equation*}\tag{S14}
    |\bar{\psi}^{R}_\pm\rangle=\frac{1}{\sqrt{\langle\psi^R_\pm|\psi^R_\pm\rangle}}|\psi^R_\pm\rangle
\end{equation*}
If we then define the RR Berry connection using $|\bar{\psi}^{R}_\pm\rangle$, it can be simply computed as
\begin{equation*}\tag{S15}
    \begin{split}
        \langle\bar{\psi}^R_\pm|i\nabla_\mathbf{k}\bar{\psi}^R_\pm\rangle=&\frac{1}{\sqrt{\langle\psi^R_\pm|\psi^R_\pm\rangle}}\langle\psi^R_\pm|\left(i\nabla_\mathbf{k}\frac{1}{\sqrt{\langle\psi^R_\pm|\psi^R_\pm\rangle}}|\psi^R_\pm\rangle\right)\\
        =&\frac{\langle\psi^R_\pm|i\nabla_\mathbf{k}\psi^R_\pm\rangle}{\langle\psi^R_\pm|\psi^R_\pm\rangle}+\sqrt{\langle\psi^R_\pm|\psi^R_\pm\rangle}i\nabla_\mathbf{k}\left(\frac{1}{\sqrt{\langle\psi^R_\pm|\psi^R_\pm\rangle}}\right)\\
        =&\frac{\langle\psi^R_\pm|i\nabla_\mathbf{k}\psi^R_\pm\rangle}{\langle\psi^R_\pm|\psi^R_\pm\rangle}-\frac{1}{2}\frac{\langle\psi^R_\pm|i\nabla_\mathbf{k}\psi^R_\pm\rangle}{\langle\psi^R_\pm|\psi^R_\pm\rangle}+\frac{1}{2}\frac{\langle-i\nabla_\mathbf{k}\psi^R_\pm|\psi^R_\pm\rangle}{\langle\psi^R_\pm|\psi^R_\pm\rangle}\\
        =&\mathbf{A}^{RR}_\pm-\frac{1}{2}\left(\mathbf{A}^{RR}_\pm-(\mathbf{A}^{RR}_\pm)^*\right)\\
        =&\operatorname{Re}\mathbf{A}^{RR}_\pm
    \end{split}
\end{equation*}
Since $|\bar{\psi}^R_\pm\rangle$ only gives the real part of the RR Berry connections, yet both real and imaginary parts of the Berry connection appear in the equation of motion, we need to compute the RR Berry connection using $|\psi^R_\pm\rangle$ instead. An example distribution of the Berry connection in momentum space for the exciton-polariton system with $\Delta=0.008$eV is presented in Fig.~\ref{fig: wpa}.

\bibliography{refs}